\documentclass[longauth]{aa}

\usepackage{natbib}
\bibpunct{(}{)}{;}{a}{}{,} 
\usepackage{amssymb}
\usepackage{url}
\usepackage{graphicx}
\usepackage{longtable}
\usepackage[usenames,dvipsnames]{color}
\usepackage{booktabs}
\usepackage{lscape}
\newcommand{\urltt}[1]{\texttt{#1}}
\newcommand{\micron}{\mbox{$\mu$m}}
\newcommand{\msun}{\mbox{$M_\odot$}}

\newcommand{\lsun}{\mbox{$L_\odot$}}
\newcommand{\msunyr}{\mbox{\msun\ yr$^{-1}$}}
\newcommand{\hi}{\ion{H}{i}}
\newcommand{\mhi}{\mbox{$M_{\rm H{\sc I}}$}}
\newcommand{\cii}{[\ion{C}{ii}]}
\newcommand{\oi}{[\ion{O}{i}]}
\newcommand{\oii}{[\ion{O}{ii}]}
\newcommand{\lp}{\mbox{$L'$}}
\newcommand{\kms}{\mbox{km\,s$^{-1}$}}
\newcommand{\Kkmspc}{\mbox{K\,km\,s$^{-1}$\,pc$^2$}}
\newcommand{\htwo}{\mbox{H$_2$}}
\newcommand{\mhtwo}{\mbox{$M_{\rm H_2}$}}
\newcommand{\grb}{GRB\,980425}
\newcommand{\metoh}{12+\log(\mbox{O}/\mbox{H})}

\begin{document}

 \title{GRB 980425 host: [$\mathrm{C\,II}$], [$\mathrm{O\,I}$] and CO lines reveal  recent enhancement of star formation due to atomic gas inflow%
 \thanks{{\it Herschel} is an ESA space observatory with science instruments provided by European-led Principal Investigator consortia and with important participation from NASA.}
 }
 
\titlerunning{GRB 980425 host: [$\mathrm{C\,II}$], [$\mathrm{O\,I}$] and CO}
\authorrunning{Micha{\l}owski et al.}

\author{Micha{\l}~J.~Micha{\l}owski\inst{\ref{inst:roe}}
\and
J.~M.~Castro~Cer\'{o}n\inst{\ref{inst:jm}} 
\and
J.~L.~Wardlow\inst{\ref{inst:dark},\ref{inst:durham}}
\and
A.~Karska\inst{\ref{inst:tor},\ref{inst:amu}}
\and
H.~Messias\inst{\ref{inst:port}}
\and
P.~van der Werf\inst{\ref{inst:vdw}}
\and
L.~K.~Hunt\inst{\ref{inst:hunt}}
\and
M.~Baes\inst{\ref{inst:gent}}
\and
A.~J.~Castro-Tirado\inst{\ref{inst:iaa}}
G.~Gentile\inst{\ref{inst:gent},\ref{inst:brus}}
\and
J.~Hjorth\inst{\ref{inst:dark}}
\and
E.~Le Floc'h\inst{\ref{inst:sacley}} 
R.~P\'erez-Mart\'inez\inst{\ref{inst:xmmesa}}
\and
A.~Nicuesa Guelbenzu\inst{\ref{inst:taut}} 	
\and
J.~Rasmussen\inst{\ref{inst:dark},\ref{inst:dtu}}
\and
J.~R.~Rizzo\inst{\ref{inst:csic}}
\and
A.~Rossi\inst{\ref{inst:pal}}
\and 
M.~S\'anchez-Portal\inst{\ref{inst:jm}}
\and
P.~Schady\inst{\ref{inst:mpe}}
\and
J.~Sollerman\inst{\ref{inst:soll}}
\and
D.~Xu\inst{\ref{inst:xu}} 
	}

\institute{
SUPA\thanks{Scottish Universities Physics Alliance}, Institute for Astronomy, University of Edinburgh, Royal Observatory, Blakford Hill, Edinburgh, EH9 3HJ, UK, {\tt mm@roe.ac.uk} 
\label{inst:roe}
\and
Herschel Science Centre (ESA-ESAC), E-28.692 Villanueva de la Ca\~nada (Madrid), Spain \label{inst:jm}
\and
Dark Cosmology Centre, Niels Bohr Institute, University of Copenhagen, Juliane Maries Vej 30, DK-2100 Copenhagen \O, Denmark  \label{inst:dark}
\and
Centre for Extragalactic Astronomy, Department of Physics, Durham University, South Road, Durham DH13LE, UK \label{inst:durham}
\and
Centre for Astronomy, Nicolaus Copernicus University, Faculty of Physics, Astronomy and Informatics, Grudziadzka 5, PL-87100 Toru\'{n}, Poland \label{inst:tor}
\and
Astronomical Observatory, Adam Mickiewicz University, S{\l}oneczna 36, 60-268, Pozna\'{n}, Poland  \label{inst:amu}
\and
Centro de Astronomia e Astrof\'{i}sica da Universidade de Lisboa,  Observat\'{o}rio Astron\'{o}mico de Lisboa, Tapada da Ajuda, 1349-018 Lisbon, Portugal \label{inst:port}
\and
Leiden Observatory, Leiden University, P.O. Box 9513, NL-2300 RA Leiden, The Netherlands \label{inst:vdw}
\and
INAF-Osservatorio Astrofisico di Arcetri, Largo E. Fermi 5, I-50125 Firenze, Italy \label{inst:hunt}
\and
Sterrenkundig Observatorium, Universiteit Gent, Krijgslaan 281-S9, 9000, Gent, Belgium  \label{inst:gent}
\and
Instituto de Astrof\'\i sica de Andaluc\'\i a (IAA-CSIC), Glorieta de la Astronom\'{\i}a s/n, E-18.008 Granada, Spain \label{inst:iaa} 
\and
Department of Physics and Astrophysics, Vrije Universiteit Brussel, Pleinlaan 2, 1050 Brussels, Belgium \label{inst:brus}
\and
Laboratoire AIM-Paris-Saclay, CEA/DSM/Irfu - CNRS - Universit\'e Paris Diderot, CE-Saclay, pt courrier 131, F-91191 Gif-sur-Yvette, France \label{inst:sacley}
\and
XMM Science Operation Centre, ESAC/ESA, Villafranca del Castillo, Spain; INSA S.A. \label{inst:xmmesa}
\and
Th\"uringer Landessternwarte Tautenburg, Sternwarte 5, D-07778 Tautenburg, Germany \label{inst:taut}
\and
Technical University of Denmark, Department of Physics, Fysikvej, building 309, DK-2800 Kgs. Lyngby, Denmark \label{inst:dtu}
\and
Centro de Astrobiolog\'{\i}a (INTA-CSIC), Ctra. M-108, km.~4, E-28850 Torrej\'on de Ardoz, Madrid, Spain \label{inst:csic}
\and           
INAF-IASF Bologna, Via Gobetti 101, I-40129 Bologna, Italy \label{inst:pal}
\and
Max-Planck-Institut f{\"u}r Extraterrestrische Physik, Giessenbachstra\ss e, 85748, Garching, Germany \label{inst:mpe}
\and
The Oskar Klein Centre, Department of Astronomy, AlbaNova, Stockholm University, 106 91 Stockholm, Sweden\label{inst:soll} 
\and
Key Laboratory of Space Astronomy and Technology, National Astronomical Observatories, Chinese Academy of Sciences, 20A Datun Road, Beijing 100012, China \label{inst:xu}
}

\abstract{
Accretion of gas from the intergalactic medium is required to fuel star formation in galaxies. We have recently suggested that this process can be studied using host galaxies of gamma-ray bursts (GRBs).}
{Our aim is to test this possibility by studying in detail the properties of gas in the closest galaxy hosting a GRB (980425). }
{We obtained the first ever far-infrared (FIR) line observations of a GRB host, namely {\it Herschel}/PACS resolved {\cii} $158\,\micron$ and {\oi} $63\,\micron$ spectroscopy, as well as APEX/SHeFI CO(2-1) line detection and ALMA CO(1-0) observations of the GRB\,980425 host.}
{The GRB\,980425 host  has elevated {\cii}/FIR and {\oi}/FIR ratios and higher values of star formation rate (SFR) derived from line ({\cii}, {\oi}, H$\alpha$) than from continuum (UV, IR, radio) indicators. {\cii} emission exhibits a normal morphology, peaking at the galaxy center, whereas {\oi} is concentrated close to the GRB position and the nearby Wolf-Rayet region. The high {\oi} flux indicates high radiation field and gas density at these positions, as derived from Photo Dissociation Region modelling. The {\cii}/CO luminosity ratio of the GRB\,980425 host is close to the highest values found for local star-forming galaxies. Indeed, its CO-derived molecular gas mass is low given its SFR and metallicity, but the {\cii}-derived molecular gas mass is close to the expected value.}
{The {\oi} and {\hi} concentrations as well as the high radiation field and density close to the GRB position are consistent with the hypothesis of a  very recent (at most a few tens of Myr ago) inflow of atomic gas triggering star formation.
In this scenario dust has not had time to build up (explaining high line-to-continuum ratios). Such a recent enhancement of star-formation activity would indeed manifest itself in high $\mbox{SFR}_{\rm line}/\mbox{SFR}_{\rm continuum}$ ratios, because the line indicators are sensitive only to recent ($\lesssim10\,$Myr) activity, whereas the continuum indicators measure the  SFR averaged over much longer periods ($\sim100\,$Myr). Within a sample of 32 other GRB hosts, 20 exhibit $\mbox{SFR}_{\rm line}/\mbox{SFR}_{\rm continuum}>1$, with a mean ratio of $1.74\pm0.32$. 
This is consistent with a very recent enhancement of star formation being common
among GRB hosts, so galaxies which have recently experienced  inflow of gas may preferentially host stars exploding as GRBs. Therefore GRBs may be used to select unique samples of galaxies suitable for the investigation of recent gas accretion.}


\maketitle

\section{Introduction}
\label{sec:intro}

\begin{figure}
\begin{center}
\includegraphics[width=0.5\textwidth,viewport= 40 10 400 400,clip]{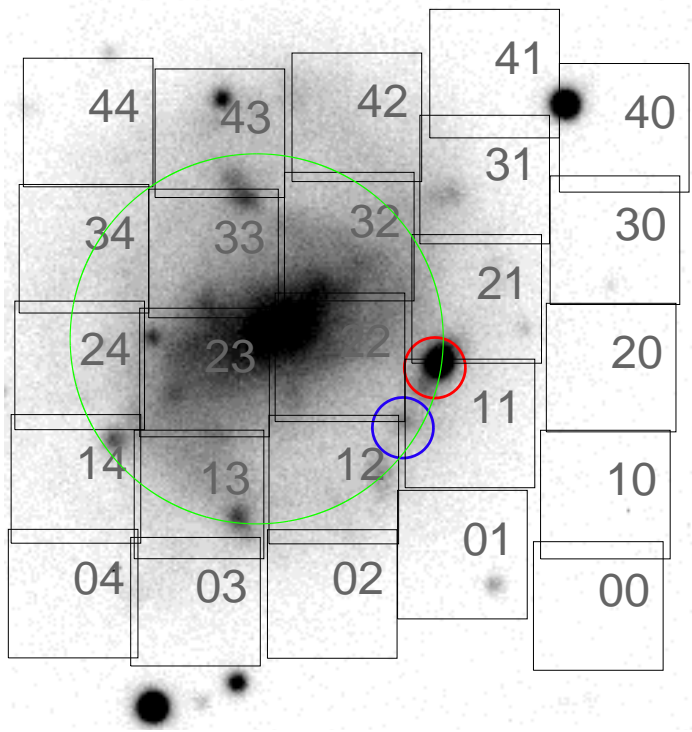} 
\end{center}
\caption{$R$-band image of the GRB 980425 host \citep{sollerman05}  with the position and numbers of the PACS spaxels indicated.
The panel is  $60''\times60''$  ($11\mbox{ kpc}\times11\mbox{ kpc}$). The {\it blue} and {\it red circles}  show the position of the GRB and the WR region, respectively. 
{\it Green circle} shows the APEX beam for the CO(2-1) observations.
}
\label{fig:thumb}
\end{figure}

One of the most important aspects of the evolution of the Universe is how galaxies acquire gas which fuels star formation. 
Numerical galaxy-formation models require significant gas inflows from the intergalactic medium (IGM) to fuel star formation  \citep[e.g.][]{schaye10}, and indeed the current gas reservoirs in many galaxies are too low to sustain the current level of star formation, even for normal galaxies like the Milky Way \citep[e.g.][]{draine09}.
However, despite many indirect evidence for gas inflows \citep[e.g.][]{sancisi08,sanchezalmeida13,sanchezalmeida14,sanchezalmeida14b,stott13,wang15}, they have been claimed to be observationally detected in only a handful of galaxies \citep{ribaudo11,martin14,michalowski15hi,turner15,rauch16}, including host galaxies of long ($\mbox{duration}>2$~s) gamma-ray bursts (GRBs). 

GRBs are explosions of very massive and short-lived stars 
(e.g.~\citealt{hjorthnature,stanek}; for a review see \citealt{hjorthsn}), so they  pinpoint locations of recent star formation.
Star formation is usually assumed to be fuelled by molecular gas \citep{carilli13,rafelski16}, but
several GRB host galaxies 
show a deficit in molecular gas \citep[\htwo;][]{hatsukade14,stanway15}, 
which is unusual for galaxies with normal star formation rates  (SFRs; unlike for extreme starbursts).
 This deficiency is not due to a high CO-to-{\htwo} conversion factor \citep[which happens at low metallicity;][]{bolatto13}, as CO-targeted GRB hosts have  
  metallicities $\metoh\sim8.7$--$9.0$  \citep{castrotirado07, levesque10b,stanway15b}, using the calibrations of \citet{pagel79}, \citet{kewley02}, \citet{pettini04}, and \citet{maiolino08}, close to the solar metallicity of $\metoh\sim8.66$ \citep{asplund04}.
 Moreover,  optical spectroscopy of GRB afterglows implies that the molecular phase constitutes only a small fraction of the gas along the GRB line-of-sight
 \citep{vreeswijk04,fynbo06b,tumlinson07,prochaska09,delia10,delia14,kruhler13,friis15}.

On the other hand the Australia Telescope Compact Array (ATCA) 21\,cm line survey of GRB host galaxies revealed high levels of  atomic hydrogen ({\hi}),
suggesting that the connection between atomic gas and star formation is stronger than previously thought \citep{michalowski15hi}. Star formation may
directly be fuelled by atomic gas, as has been theoretically shown to be possible  \citep{glover12,krumholz12,hu16},
and supported by the existence of {\hi}-dominated star-forming regions in other galaxies \citep{bigiel08,bigiel10, fumagalli08,elmegreen16}.
This can happen in a recently-acquired low metallicity gas (even if the metallicity in other parts of a galaxy is higher) near the onset of star formation, because cooling of gas (necessary for star formation) is faster than the {\hi}-to-{\htwo} conversion 
\citep{krumholz12}. 
Indeed, large atomic gas reservoirs, together with low molecular gas masses 
\citep{hatsukade14,stanway15} 
and stellar masses 
\citep{perley13,perley15,vergani15}, 
indicate that GRB hosts are preferentially galaxies which have very recently started a star formation episode,
providing a natural route for forming GRBs in low-metallicity environments, as found for most GRB hosts
 \citep{fruchter06,modjaz08, levesque10c, han10, boissier13,schulze15,vergani15,japelj16,perley16b}, except of a few examples of 
 hosts with solar or super-solar metallicities \citep{prochaska09,levesque10b,kruhler12,savaglio12,elliott13,schulze14, hashimoto15,schady15,stanway15b}. 
Indeed,  the GRB collapsar model requires that most of the GRB progenitors have low metallicity (below solar) in order to reduce the loss of mass  and angular momentum 
\citep[required for launching the jet;][]{yoon05,yoon06, woosley06}. 
We note however, that other models, while still predicting the metallicity preference \citep[e.g.][]{izzard04,podsiadlowski04,detmers08}, do allow higher metallicities thanks to  differential rotation \citep{georgy12}, binary evolution \citep{podsiadlowski10,vandenheuvel13}, or weaker magnetic fields \citep{petrovic05}.

Summarising the ATCA {\hi} data support a scenario whereby GRBs are preferentially produced when low-metallicity gas accretes onto a galaxy and undergoes rapid cooling and star formation before it either forms {\htwo} or mixes with the higher-metallicity gas in the remainder of the galaxy.
This scenario provides a natural explanation for  the low-metallicity and low-{\mhtwo} preferences.
In contrast, at later stages of star formation molecular gas is the dominant phase in the interstellar medium, but the metals are well mixed, and gas has been further enriched, so massive stars do not end their lives as GRBs, and such metal- and molecular-rich galaxies do not become GRB hosts.

The gas inflow scenario is also supported by the existence of the companion {\hi} object with no optical counterpart $\sim19$\,kpc from the GRB\,060505 host, which may be a stream of gas inflowing on this galaxy, and  by the fact that the {\hi} centroids of the GRB\,980425 and 060505 hosts do not coincide with the optical centres of these galaxies, but are located close to the GRB positions \citep{michalowski15hi}. The concentration of {\hi} close to the GRB\,980425 position has been confirmed with high-resolution {\hi} imagining by the Giant Metrewave Radio Telescope \citep[GMRT;][]{arabsalmani15b}.

\begin{figure*}
\begin{center}
\begin{tabular}{cc}
\includegraphics[width=0.48\textwidth]{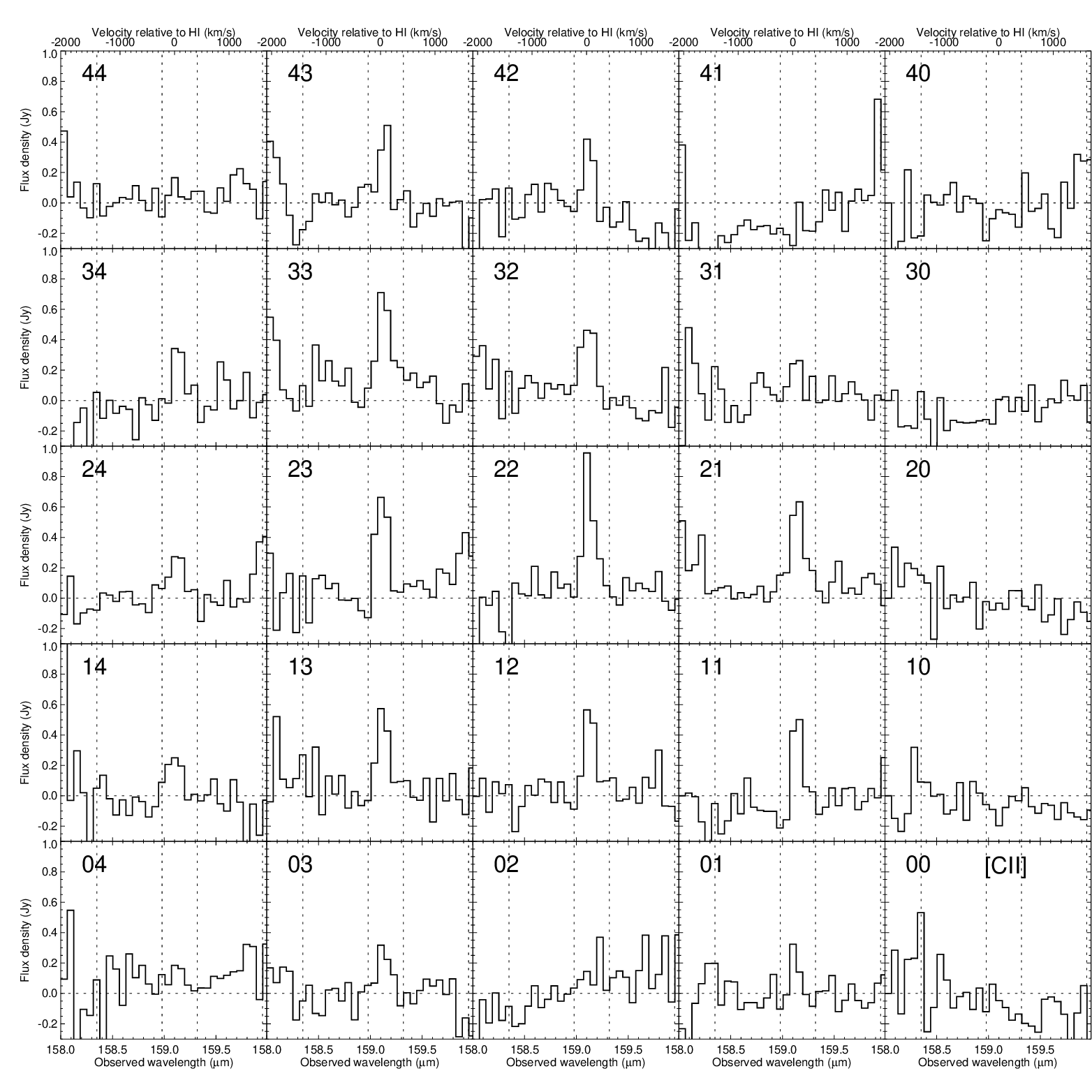} & \includegraphics[width=0.48\textwidth]{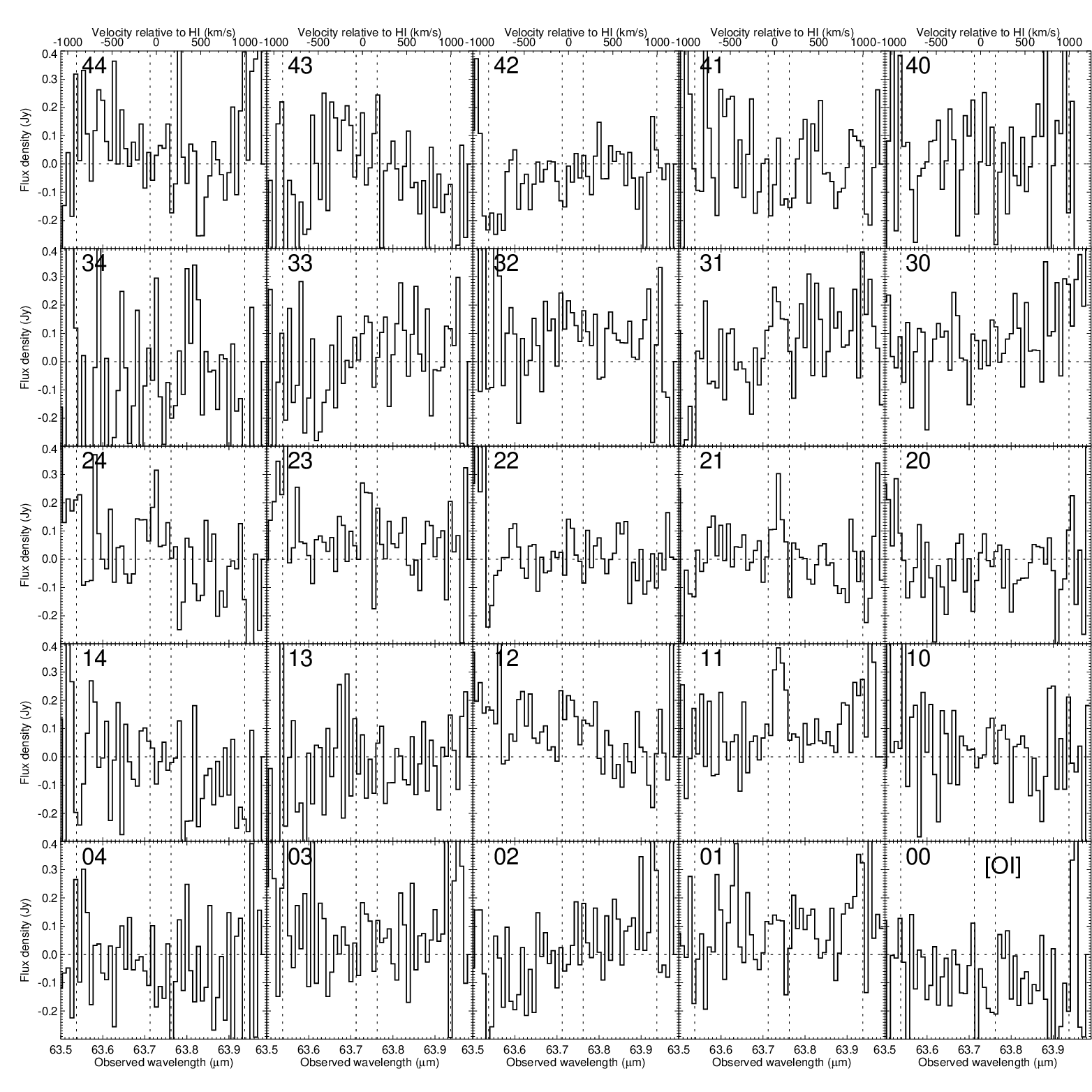} \\
\includegraphics[width=0.44\textwidth,clip]{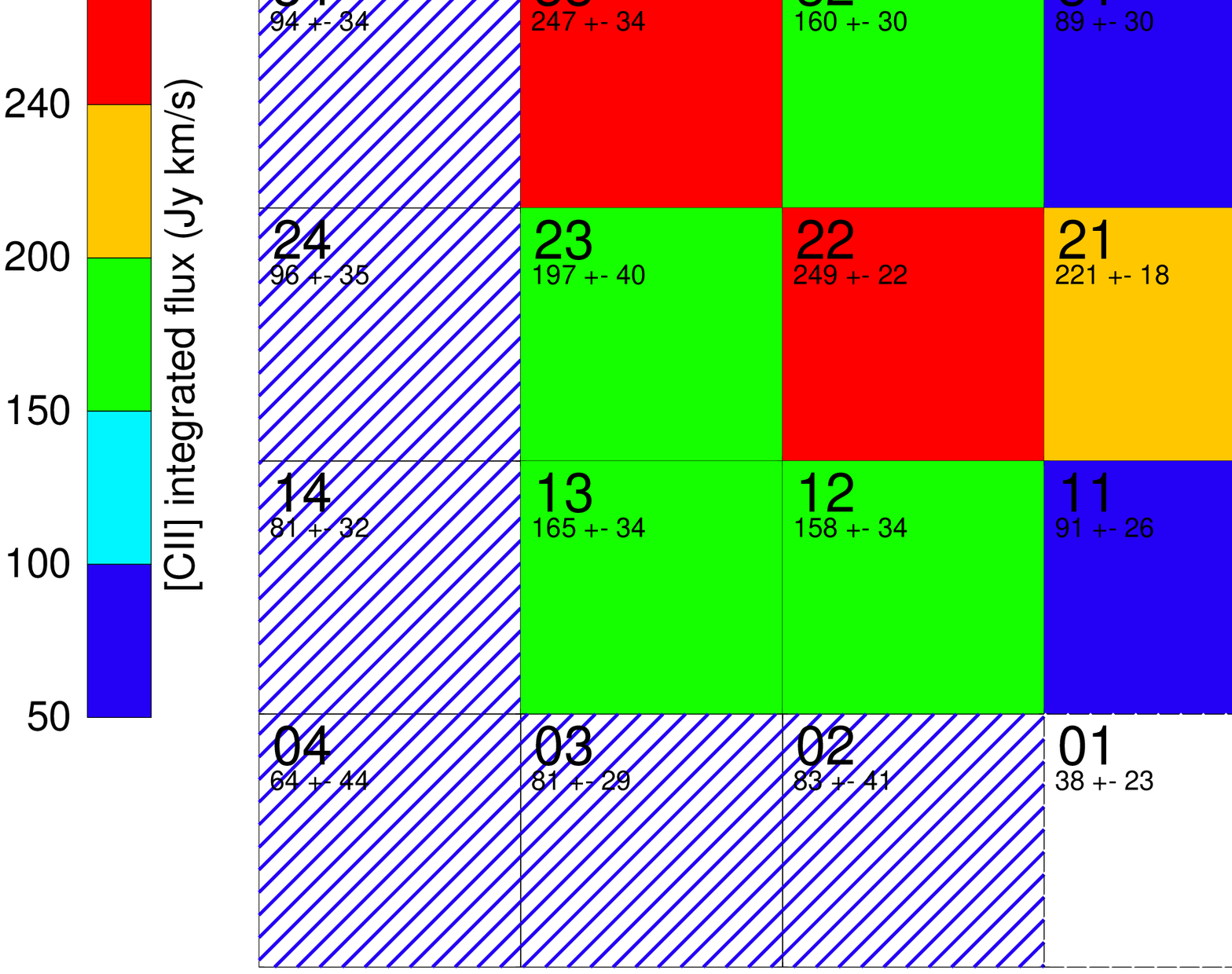} & \includegraphics[width=0.44\textwidth,clip]{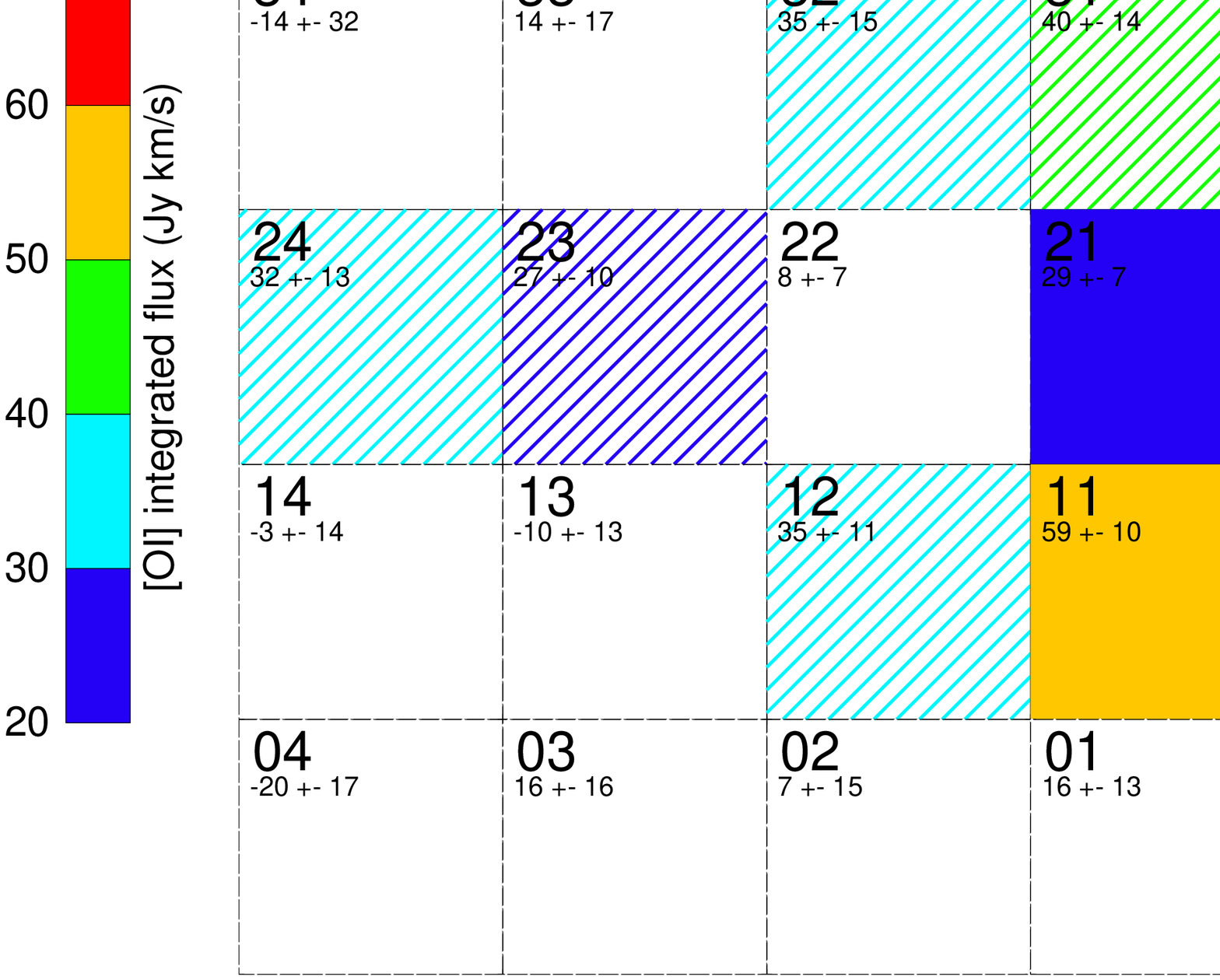}\\
\includegraphics[width=0.44\textwidth,clip]{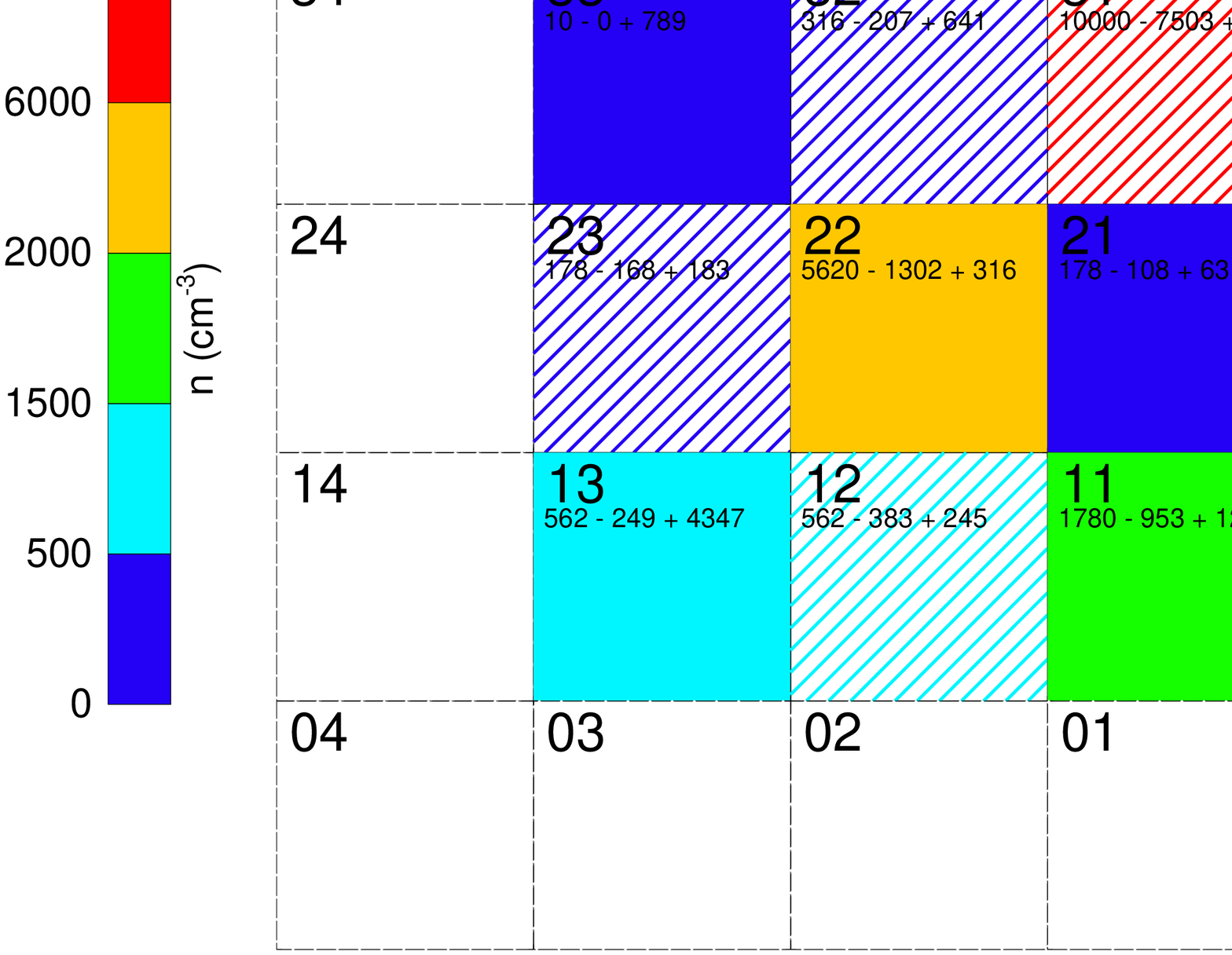} & \includegraphics[width=0.44\textwidth,clip]{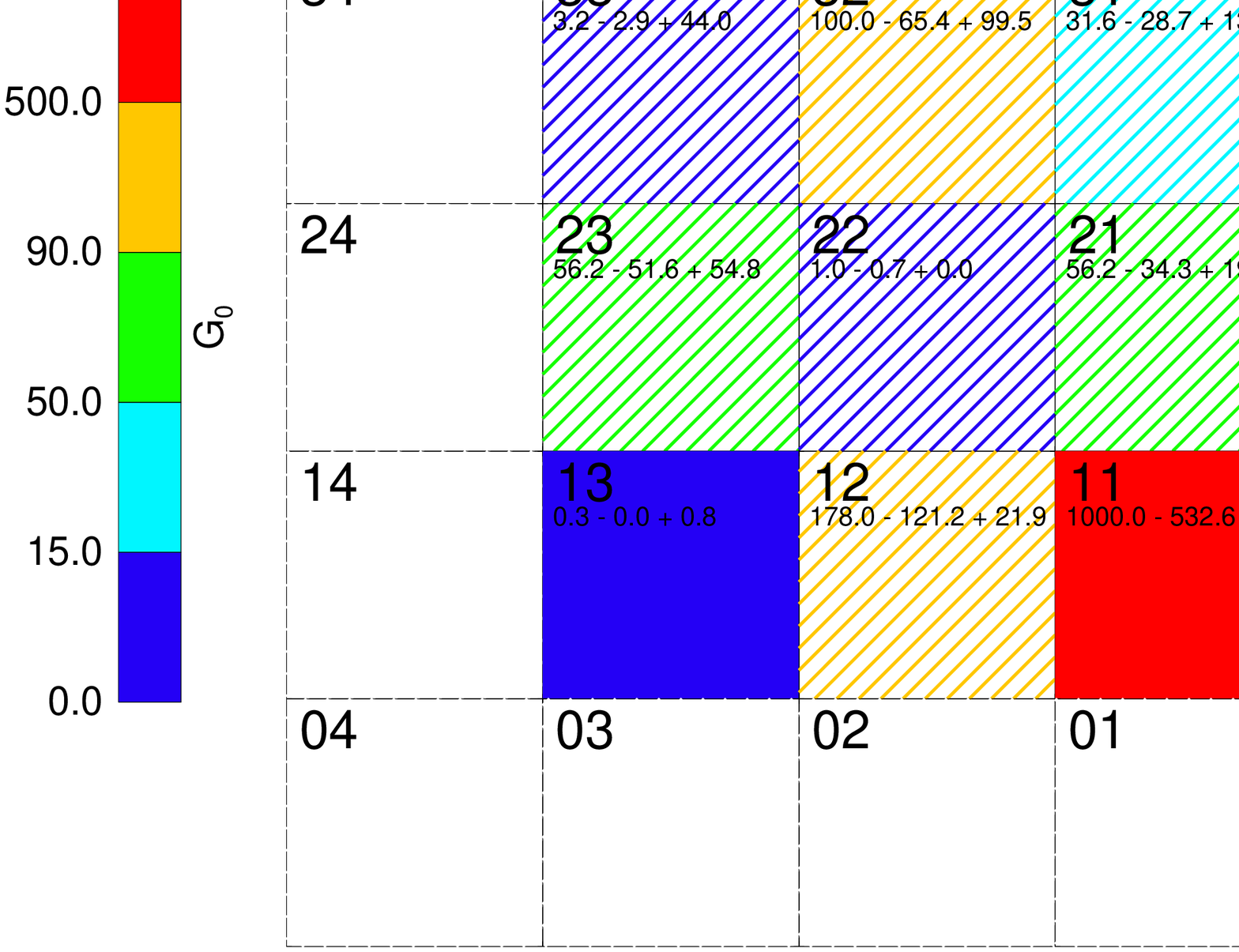}
\end{tabular}
\end{center}
\caption{{\it Top}: 
 {\cii} ({\it left}) and {\oi} ({\it right}) spectra of each spaxel. The integrated fluxes were measured within inner dotted lines, and the errors within the outer lines excluding the inner region.
{\it Middle}: {\cii} ({\it left}) and {\oi} ({\it right}) spatial distribution of integrated flux in each spaxel. 
{\it Bottom}: distribution of derived density and radiation field strength.
In order to improve the readability of the plots we adopted the following convention. The line-filled boxes correspond to upper limits on the corresponding property if the flux measurement has $<3\sigma$ significance, but if the measured value is below the lowest value on the colour bar, then the box is white. If the property cannot be measured ($n$ and $G_0$ for spaxels with no detections at {\cii} and {\oi}), then the box is also white with no values indicated. 
The {\cii} emission exhibits a normal radial profile, whereas  {\oi} emission is concentrated close to the WR region.
}
\label{fig:spax}
\end{figure*}

\begin{landscape}
\begin{table}
\tiny
\centering
\caption{Line and continuum fluxes and luminosities of each spaxel and the entire host.\label{tab:flux}}
\medskip
\begin{tabular}{lrrrrrrrrrrrr}
\hline\hline
       & \multicolumn{5}{c}{\cii} & \multicolumn{5}{c}{\oi} \\
\cmidrule(r){2-6} \cmidrule(l){7-11}
Reg & \multicolumn{1}{c}{$F_{\rm int}$} & S/N & \multicolumn{1}{c}{$F_{\rm int}$} & \multicolumn{1}{c}{$L$} & \multicolumn{1}{c}{$\lp$} & \multicolumn{1}{c}{$F_{\rm int}$} & S/N & \multicolumn{1}{c}{$F_{\rm int}$} & \multicolumn{1}{c}{$L$} & \multicolumn{1}{c}{$\lp$} & \multicolumn{1}{c}{$F_{100\,\micron}$} & \multicolumn{1}{c}{$L_{\rm FIR}$}  \\
       & \multicolumn{1}{c}{(Jy km s$^{-1}$)} &  & \multicolumn{1}{c}{($10^{-17}$W m$^{-2}$)}    & \multicolumn{1}{c}{($10^{5}L_\odot$)} & \multicolumn{1}{c}{($10^{6}\mbox{K km s}^{-1} \mbox{ pc}^2$)} & \multicolumn{1}{c}{(Jy km s$^{-1}$)} &  & \multicolumn{1}{c}{($10^{-17}$W m$^{-2}$)}    & \multicolumn{1}{c}{($10^{5}L_\odot$)} & \multicolumn{1}{c}{($10^{6}\mbox{K km s}^{-1} \mbox{ pc}^2$)} & \multicolumn{1}{c}{(mJy)} & \multicolumn{1}{c}{($10^6\lsun$)}  \\
(1)    & \multicolumn{1}{c}{(2)}     & (3)  & \multicolumn{1}{c}{(4)}           & \multicolumn{1}{c}{(5)} & \multicolumn{1}{c}{(6)} & \multicolumn{1}{c}{(7)} & (8) & \multicolumn{1}{c}{(9)} & \multicolumn{1}{c}{(10)} & \multicolumn{1}{c}{(11)} & \multicolumn{1}{c}{(12)} & \multicolumn{1}{c}{(13)}  \\
\hline
Host & $2073\pm129$ &$16.1$ &$13.14\pm0.82$ &$55.9\pm3.5$ &$25.46\pm1.58$ &$262\pm49$ &$5.3$ &$4.15\pm0.78$ &$17.7\pm3.3$ &$0.52\pm0.10$ &$327\pm12$ &$ 483.2\pm5.6$ \\
44 & $ 37\pm23$ & $ 1.6$ & $ 0.24\pm0.15$ & $ 1.0\pm0.6$ & $ 0.46\pm0.29$ & $ 8\pm17$ & $ 0.5$ & $ 0.12\pm0.27$ & $ 0.5\pm1.1$ & $ 0.02\pm0.03$ & $ 9\pm\phantom{1}3$ & $ 20.3\pm6.6$ \\
43 & $ 115\pm26$ & $ 4.4$ & $ 0.73\pm0.17$ & $ 3.1\pm0.7$ & $ 1.41\pm0.32$ & $ 13\pm17$ & $ 0.8$ & $ 0.21\pm0.28$ & $ 0.9\pm1.2$ & $ 0.03\pm0.03$ & $ 20\pm\phantom{1}3$ & $ 45.7\pm6.6$ \\
42 & $ 73\pm41$ & $ 1.8$ & $ 0.46\pm0.26$ & $ 2.0\pm1.1$ & $ 0.89\pm0.50$ & $ -11\pm11$ & $ -1.0$ & $ -0.17\pm0.17$ & $ -0.7\pm0.7$ & $ -0.02\pm0.02$ & $ 8\pm\phantom{1}3$ & $ 17.4\pm6.6$ \\
41 & $ -110\pm63$ & $ -1.7$ & $ -0.70\pm0.40$ & $ -3.0\pm1.7$ & $ -1.35\pm0.78$ & $ -21\pm14$ & $ -1.5$ & $ -0.33\pm0.22$ & $ -1.4\pm1.0$ & $ -0.04\pm0.03$ & $ -3\pm\phantom{1}3$ & $ -7.2\pm6.6$ \\
40 & $ -66\pm41$ & $ -1.6$ & $ -0.42\pm0.26$ & $ -1.8\pm1.1$ & $ -0.81\pm0.51$ & $ 6\pm20$ & $ 0.3$ & $ 0.10\pm0.32$ & $ 0.4\pm1.4$ & $ 0.01\pm0.04$ & $ -1\pm\phantom{1}3$ & $ -1.2\pm6.6$ \\
34 & $ 94\pm34$ & $ 2.7$ & $ 0.60\pm0.22$ & $ 2.5\pm0.9$ & $ 1.16\pm0.42$ & $ -14\pm32$ & $ -0.4$ & $ -0.22\pm0.50$ & $ -0.9\pm2.1$ & $ -0.03\pm0.06$ & $ 20\pm\phantom{1}3$ & $ 45.8\pm6.6$ \\
33 & $ 247\pm34$ & $ 7.2$ & $ 1.57\pm0.22$ & $ 6.7\pm0.9$ & $ 3.04\pm0.42$ & $ 14\pm17$ & $ 0.9$ & $ 0.23\pm0.26$ & $ 1.0\pm1.1$ & $ 0.03\pm0.03$ & $ 33\pm\phantom{1}3$ & $ 75.4\pm6.6$ \\
32 & $ 160\pm30$ & $ 5.4$ & $ 1.02\pm0.19$ & $ 4.3\pm0.8$ & $ 1.97\pm0.36$ & $ 35\pm15$ & $ 2.3$ & $ 0.55\pm0.24$ & $ 2.3\pm1.0$ & $ 0.07\pm0.03$ & $ 33\pm\phantom{1}3$ & $ 74.7\pm6.6$ \\
31 & $ 89\pm30$ & $ 3.0$ & $ 0.57\pm0.19$ & $ 2.4\pm0.8$ & $ 1.10\pm0.36$ & $ 40\pm14$ & $ 2.9$ & $ 0.63\pm0.22$ & $ 2.7\pm0.9$ & $ 0.08\pm0.03$ & $ 7\pm\phantom{1}3$ & $ 16.5\pm6.6$ \\
30 & $ -28\pm37$ & $ -0.8$ & $ -0.18\pm0.23$ & $ -0.7\pm1.0$ & $ -0.34\pm0.45$ & $ 14\pm14$ & $ 1.0$ & $ 0.21\pm0.21$ & $ 0.9\pm0.9$ & $ 0.03\pm0.03$ & $ 3\pm\phantom{1}3$ & $ 7.7\pm6.6$ \\
24 & $ 96\pm35$ & $ 2.8$ & $ 0.61\pm0.22$ & $ 2.6\pm0.9$ & $ 1.18\pm0.43$ & $ 32\pm13$ & $ 2.4$ & $ 0.50\pm0.21$ & $ 2.1\pm0.9$ & $ 0.06\pm0.03$ & $ 24\pm\phantom{1}3$ & $ 55.3\pm6.6$ \\
23 & $ 197\pm40$ & $ 5.0$ & $ 1.25\pm0.25$ & $ 5.3\pm1.1$ & $ 2.41\pm0.49$ & $ 27\pm10$ & $ 2.7$ & $ 0.43\pm0.16$ & $ 1.8\pm0.7$ & $ 0.05\pm0.02$ & $ 39\pm\phantom{1}3$ & $ 90.3\pm6.6$ \\
22 & $ 249\pm22$ & $ 11.3$ & $ 1.58\pm0.14$ & $ 6.7\pm0.6$ & $ 3.06\pm0.27$ & $ 8\pm7$ & $ 1.0$ & $ 0.12\pm0.12$ & $ 0.5\pm0.5$ & $ 0.01\pm0.01$ & $ 47\pm\phantom{1}3$ & $ 108.3\pm6.6$ \\
21 & $ 221\pm18$ & $ 12.3$ & $ 1.40\pm0.11$ & $ 6.0\pm0.5$ & $ 2.71\pm0.22$ & $ 29\pm7$ & $ 4.4$ & $ 0.46\pm0.10$ & $ 1.9\pm0.4$ & $ 0.06\pm0.01$ & $ 40\pm\phantom{1}3$ & $ 90.9\pm6.6$ \\
20 & $ -12\pm34$ & $ -0.3$ & $ -0.07\pm0.21$ & $ -0.3\pm0.9$ & $ -0.14\pm0.41$ & $ -0\pm9$ & $ -0.1$ & $ -0.01\pm0.14$ & $ -0.0\pm0.6$ & $ -0.00\pm0.02$ & $ 11\pm\phantom{1}3$ & $ 24.4\pm6.6$ \\
14 & $ 81\pm32$ & $ 2.5$ & $ 0.51\pm0.20$ & $ 2.2\pm0.9$ & $ 0.99\pm0.39$ & $ -3\pm14$ & $ -0.2$ & $ -0.05\pm0.22$ & $ -0.2\pm0.9$ & $ -0.01\pm0.03$ & $ 18\pm\phantom{1}3$ & $ 42.2\pm6.6$ \\
13 & $ 165\pm34$ & $ 4.8$ & $ 1.04\pm0.22$ & $ 4.4\pm0.9$ & $ 2.02\pm0.42$ & $ -10\pm13$ & $ -0.8$ & $ -0.16\pm0.20$ & $ -0.7\pm0.9$ & $ -0.02\pm0.03$ & $ 22\pm\phantom{1}3$ & $ 50.4\pm6.6$ \\
12 & $ 158\pm34$ & $ 4.7$ & $ 1.00\pm0.21$ & $ 4.3\pm0.9$ & $ 1.94\pm0.41$ & $ 35\pm11$ & $ 3.1$ & $ 0.56\pm0.18$ & $ 2.4\pm0.8$ & $ 0.07\pm0.02$ & $ 33\pm\phantom{1}3$ & $ 76.5\pm6.6$ \\
11 & $ 91\pm26$ & $ 3.6$ & $ 0.58\pm0.16$ & $ 2.4\pm0.7$ & $ 1.11\pm0.31$ & $ 59\pm10$ & $ 6.1$ & $ 0.94\pm0.16$ & $ 4.0\pm0.7$ & $ 0.12\pm0.02$ & $ 56\pm\phantom{1}3$ & $ 127.1\pm6.6$ \\
10 & $ -51\pm24$ & $ -2.1$ & $ -0.32\pm0.15$ & $ -1.4\pm0.7$ & $ -0.62\pm0.30$ & $ 9\pm16$ & $ 0.5$ & $ 0.14\pm0.26$ & $ 0.6\pm1.1$ & $ 0.02\pm0.03$ & $ 6\pm\phantom{1}3$ & $ 13.3\pm6.6$ \\
04 & $ 64\pm44$ & $ 1.4$ & $ 0.40\pm0.28$ & $ 1.7\pm1.2$ & $ 0.78\pm0.54$ & $ -20\pm17$ & $ -1.2$ & $ -0.32\pm0.26$ & $ -1.4\pm1.1$ & $ -0.04\pm0.03$ & $ 4\pm\phantom{1}3$ & $ 8.7\pm6.6$ \\
03 & $ 81\pm29$ & $ 2.8$ & $ 0.52\pm0.19$ & $ 2.2\pm0.8$ & $ 1.00\pm0.36$ & $ 16\pm16$ & $ 0.9$ & $ 0.25\pm0.26$ & $ 1.0\pm1.1$ & $ 0.03\pm0.03$ & $ 12\pm\phantom{1}3$ & $ 28.6\pm6.6$ \\
02 & $ 83\pm41$ & $ 2.0$ & $ 0.53\pm0.26$ & $ 2.2\pm1.1$ & $ 1.02\pm0.50$ & $ 7\pm15$ & $ 0.5$ & $ 0.11\pm0.23$ & $ 0.4\pm1.0$ & $ 0.01\pm0.03$ & $ 10\pm\phantom{1}3$ & $ 23.8\pm6.6$ \\
01 & $ 38\pm23$ & $ 1.7$ & $ 0.24\pm0.15$ & $ 1.0\pm0.6$ & $ 0.47\pm0.28$ & $ 16\pm13$ & $ 1.3$ & $ 0.25\pm0.20$ & $ 1.1\pm0.9$ & $ 0.03\pm0.03$ & $ 5\pm\phantom{1}3$ & $ 10.3\pm6.6$ \\
00 & $ -52\pm43$ & $ -1.2$ & $ -0.33\pm0.27$ & $ -1.4\pm1.2$ & $ -0.64\pm0.53$ & $ -48\pm15$ & $ -3.2$ & $ -0.75\pm0.24$ & $ -3.2\pm1.0$ & $ -0.09\pm0.03$ & $ 1\pm\phantom{1}3$ & $ 1.2\pm6.6$ \\
\hline
\end{tabular}
\tablefoot{
(1) The entire host or the spaxel number. Columns (2)--(6) concern the {\cii} and (7)--(11) concern the {\oi} line. (2,7) Integrated flux within dotted lines on Fig.~\ref{fig:spax} and \ref{fig:hostspec}. (3,8) Signal-to-noise ratio of the line. (4,9) Integrated flux in W m$^{-2}$. (5,10) Line luminosity in solar luminosity. (6,11) Line luminosity using equation 3 in \citet{solomon97}. (12) Flux at $100\,\micron$. (13) Far-infrared ($40$--$120\,\micron$) luminosity.
}
\end{table}
\end{landscape}

Here, we test the gas inflow scenario by investigating in detail the properties of gas in the GRB\,980425 host.
\object{GRB 980425} at redshift $z=0.0085$ \citep{tinney98} and its associated supernova \object{SN 1998bw} \citep{galamanature} is the closest GRB, and is located in the barred spiral galaxy \object{ESO 184-G82}. Hence, it is one of the few GRB hosts for which resolved gas properties can be studied.
Of particular importance is a Wolf-Rayet (WR) region $\sim800$ pc northwest of the GRB/SN position \citep{hammer06}, dominating the galaxy's emission at $24\,\micron$ \citep{lefloch},
$100\,\micron$ and radio \citep{michalowski14}.  This region is young \citep[$1$--$6$ Myr;][]{hammer06,christensen08}, and  exhibits the lowest metallicity among star-forming regions  within the host with $12 + \log(\mbox{O}/\mbox{H})=8.16$, i.e. $0.3$ solar \citep{christensen08}, using the calibration of \citet{pettini04} compared with $12 + \log(\mbox{O}/\mbox{H})=8.6$, i.e. $0.8$ solar \citep{sollerman05} for the entire host, using the calibration of \citet{kewley02}. 

We use a cosmological model with $H_0=70$ km s$^{-1}$ Mpc$^{-1}$,  $\Omega_\Lambda=0.7$, and $\Omega_m=0.3$, so GRB 980425 at $z=0.0085$ is at a luminosity distance of 36.5 Mpc and $1''$ corresponds to 175 pc at its redshift.

\section{Data}
\label{sec:data}

We obtained {\it Herschel} \citep{herschel} observations of the GRB 980425 host (project no.~OT2\_jmcastro\_3, PI: J.~M.~Castro~Cer\'{o}n) using the Photodetector Array Camera and Spectrometer  \citep[PACS;][]{pacs} 
with a total integration time of 4097\,s, on 20 Apr 2013\footnote{OBSIDs: 1342270641}. 
The data were taken in the line spectroscopy mode with medium chopping/nodding to off source (and off-galaxy) positions, which aids background subtraction. Due to the expected relative brightnesses of the {\oi} $63\,\micron$ and {\cii} $158\,\micron$ lines, approximately 9 times longer was spent targeting the {\oi} wavelength. The positions of the spaxels are shown on the optical image of the host in Fig.~\ref{fig:thumb}.

Data reduction for PACS was performed in the Herschel Imaging Processing Environment  \citep[{\sc Hipe};][]{hipe} v12.1.0 with version 65.0 of the PACS calibration tree and uses the {\sc Ipipe} Background Normalization script for Chop/Nod Range Scan data, which is optimised for faint sources by using off-source positions in the background subtraction and flux calibration.
To avoid introducing correlated noise we set the upsample factor to 1 during flatfielding, and to minimise signal losses we also masked the spectral regions where spectral lines are expected during this process. The final spectra are binned to be Nyquist sampled at the native PACS resolution.

We also performed CO(2-1) observations of the GRB 980425 host on 
29 Aug (precipitable water vapour [pwv] of 1.7\,mm), 
12 Sep (pwv of 0.75-0.85\,mm), 
16 Sep (pwv of 1.43--1.57\,mm), 
31 Oct (pwv of 1.22--1.96\,mm), 
and 01 Nov 2015 (pwv of  0.66--0.85\,mm) 
using the Swedish Heterodyne Facility Instrument \citep[SHeFI;][]{shefi}  mounted at the Atacama Pathfinder Experiment \citep[APEX;][]{apex}  (project no.~096.D-0280 and 096.F-9302, PI: M.~Micha{\l}owski). A total of 4.92\,hr of on-source data were obtained.  The APEX-1 single-sideband (SSB)   was tuned to the observed frequency of the CO(2-1) line of 228.6\,GHz.  
At this frequency the APEX beam size is 27{\arcsec} ($\sim4.8$\,kpc at the distance of the GRB).  All observations were completed in the 
on-off pattern and the position-switching mode. The fluxes were corrected using the main beam efficiency of 0.75.
We reduced and analysed the data using the Continuum and Line Analysis Single Dish Software ({\sc Class}) package within the Grenoble Image and Line Data Analysis Software\footnote{\url{http://www.iram.fr/IRAMFR/GILDAS}} ({\sc Gildas}; \citealt{gildas}).

We performed Band 3 ALMA observations on 1 Sep 2012 (project no.~2011.0.00046.S, PI: M.~Micha{\l}owski). A total of 67.4 min of on-source data were obtained. Four $1.875$ GHz spectral windows were centred at  $100.6$,  $102.4$, $112.5$, and $114.3$ GHz. 
Twenty three antennas and baselines ranging between $24$ and $384$ m were available. Neptune, J1733-130, and J1945-552 were used as flux, bandpass, and phase calibrators, respectively. The amount of precipitable water vapour ranged between $1.8$--$2.15$\,mm. The data reduction and analysis were done using the {\sc Casa} package \citep{casa}. 
The original spectral resolution was $\sim488$\,kHz ($\sim1.3\,\kms$).
The continuum map was presented in \citet{michalowski14}, whereas here we present a data cube at the frequency of the CO(1-0) line (114.288\,GHz)
binning 10 channels resulting in  spectral resolution of $\sim4.9\,$MHz (corresponding to $\sim13\,\kms$).
The synthesised beam of the cube is $\sim1.7\,\arcsec$.

\section{Methods}
\label{sec:mod}

We obtained three estimates of SFRs. First, we calculated 100\,{\micron} fluxes within each spaxel using the {\it Herschel}/PACS map presented in \citet{michalowski14}, and converted them to total (8--1000\,micron) infrared (IR) luminosities using the  spectral energy distribution (SED) model of the WR region in the {\grb} host \citep{michalowski14}, and to SFRs using the \citet{kennicutt} conversion for the \citet{chabrier03} initial mass function (IMF) ($\mbox{SFR}/\msunyr=10^{-10} L_{\rm IR}/\lsun$). We probe close to the peak of the SED, so if we used other templates \citep[e.g.][]{silva98,iglesias07,michalowski08,michalowski10smg,michalowski10smg4} we would obtain similar results. Then, we estimated the SFRs from the {\cii} and {\oi} lines using the conversion of \citet[][their table 3, the first two rows]{delooze14}.

Using far-infrared (FIR) luminosities integrated over $40$--$120\,\micron$ we modelled the ${\oi}/{\cii}$ and $({\oi}+{\cii})/\mbox{FIR}$ ratios  using the Photo Dissociation Region (PDR) Toolbox \citep{pound08,kaufman06}\footnote{\url{dustem.astro.umd.edu/pdrt}} with the addition of our own routine to extract errors on density and radiation field intensity from two-dimensional $\chi^2$ distributions. In this way we derived the gas density, $n$, and the ultraviolet (UV) strength of the radiation field, $G_0$ in the unit  of the radiation field strength in the solar neighbourhood of $1.6\times10^{-3}\,\mbox{erg s}^{-1}\,\mbox{cm}^{-2}$ \citep[or Habing unit;][]{habing}.

We estimated the molecular gas mass from the CO(2-1) line  assuming the flux conversion $S_{\rm CO(1-0)}=0.5\times S_{\rm CO(2-1)}$  \citep[fig.~4 in][]{carilli13}, i.e.~$L'_{\rm CO(1-0)}= 2\times L'_{\rm CO(2-1)}$, assuming the Galactic CO-to-{\htwo} conversion factor $\alpha_{\rm CO}=5\msun/(\Kkmspc)$.

Finally,  we estimated the molecular gas  mass from the {\cii} line using the {\cii}/CO(1-0) luminosity ratio of around 5000, typical for star-forming galaxies  \citep{crawford85, wolfire89,stacey91,stacey10,swinbank12,rigopoulou14,gullberg15,hughes17}. As noted by \citet{neri14}, the luminosities of these lines expressed in brightness temperature units, $L'$ (\Kkmspc), are approximately equal (within a factor of two). Hence, we substituted $L'_{\rm CO(1-0)}$ with $L'_{\cii}$ and obtained molecular gas masses with $\mhtwo=\alpha_{\rm CO}L'_{\cii}$, assuming and the same CO-to-{\htwo} conversion factor as above.

\section{Results}
\label{sec:res}

\begin{figure*}
\begin{center}
\begin{tabular}{cc}
\includegraphics[width=0.5\textwidth]{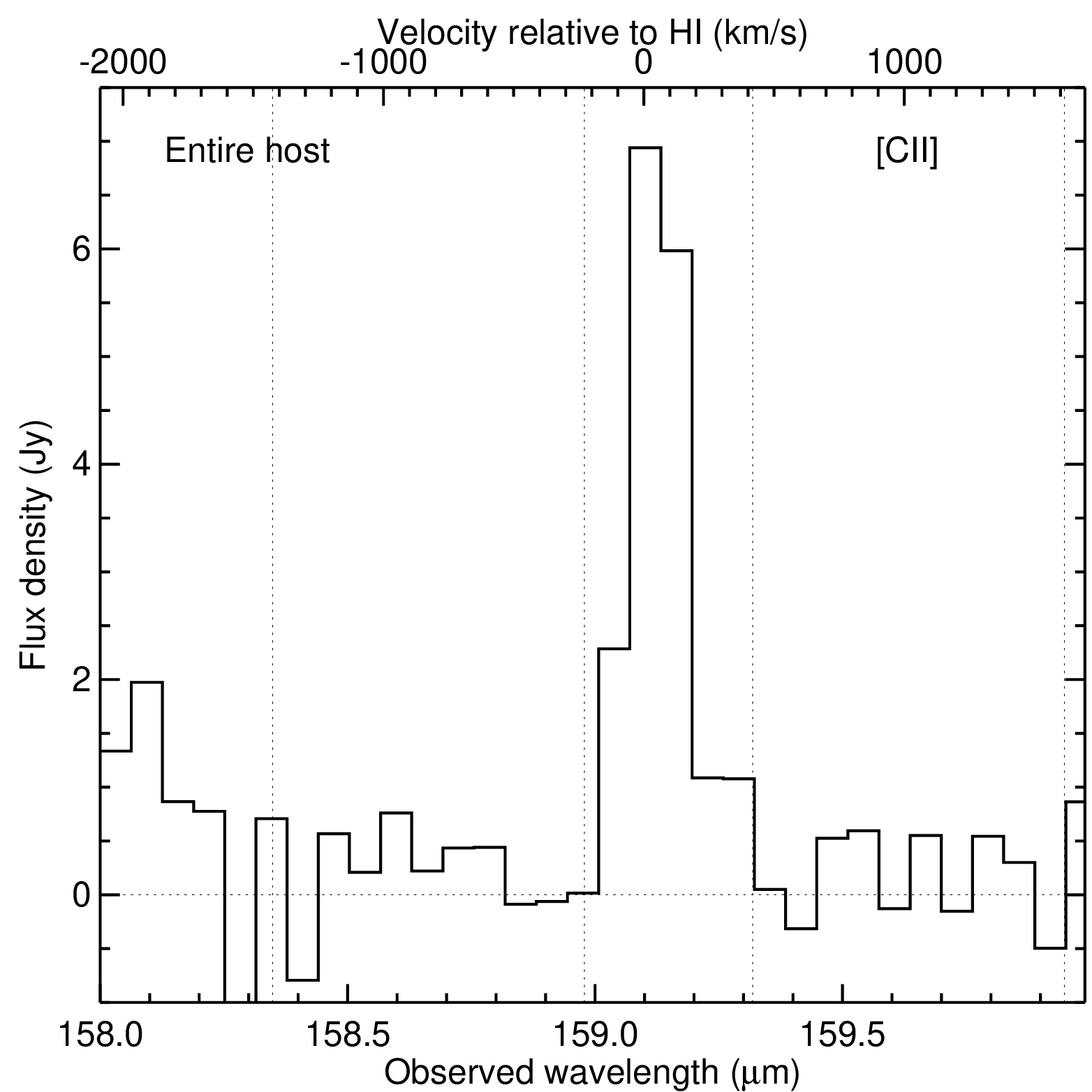} & \includegraphics[width=0.5\textwidth]{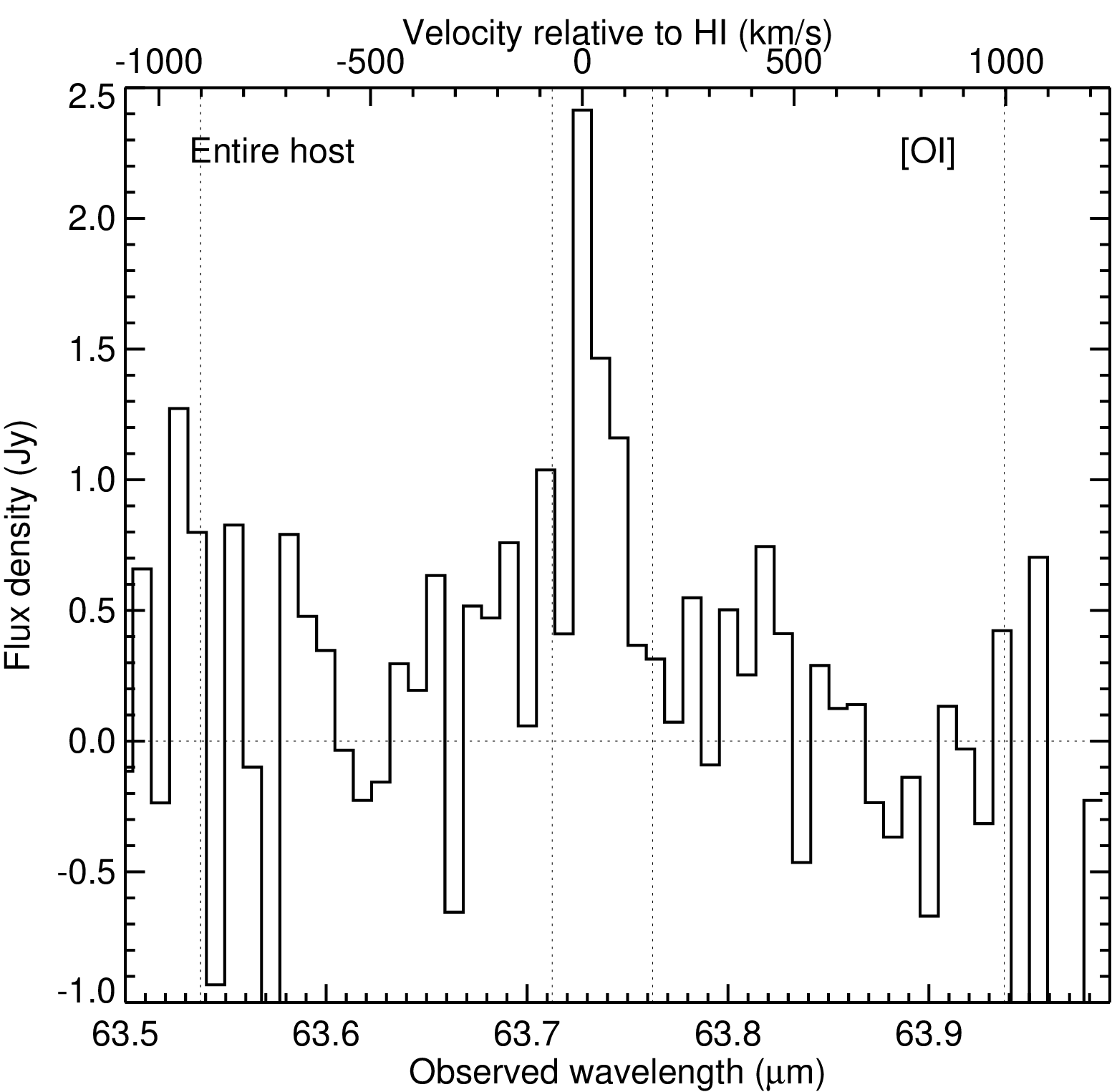} \\
{\includegraphics[width=0.5\textwidth]{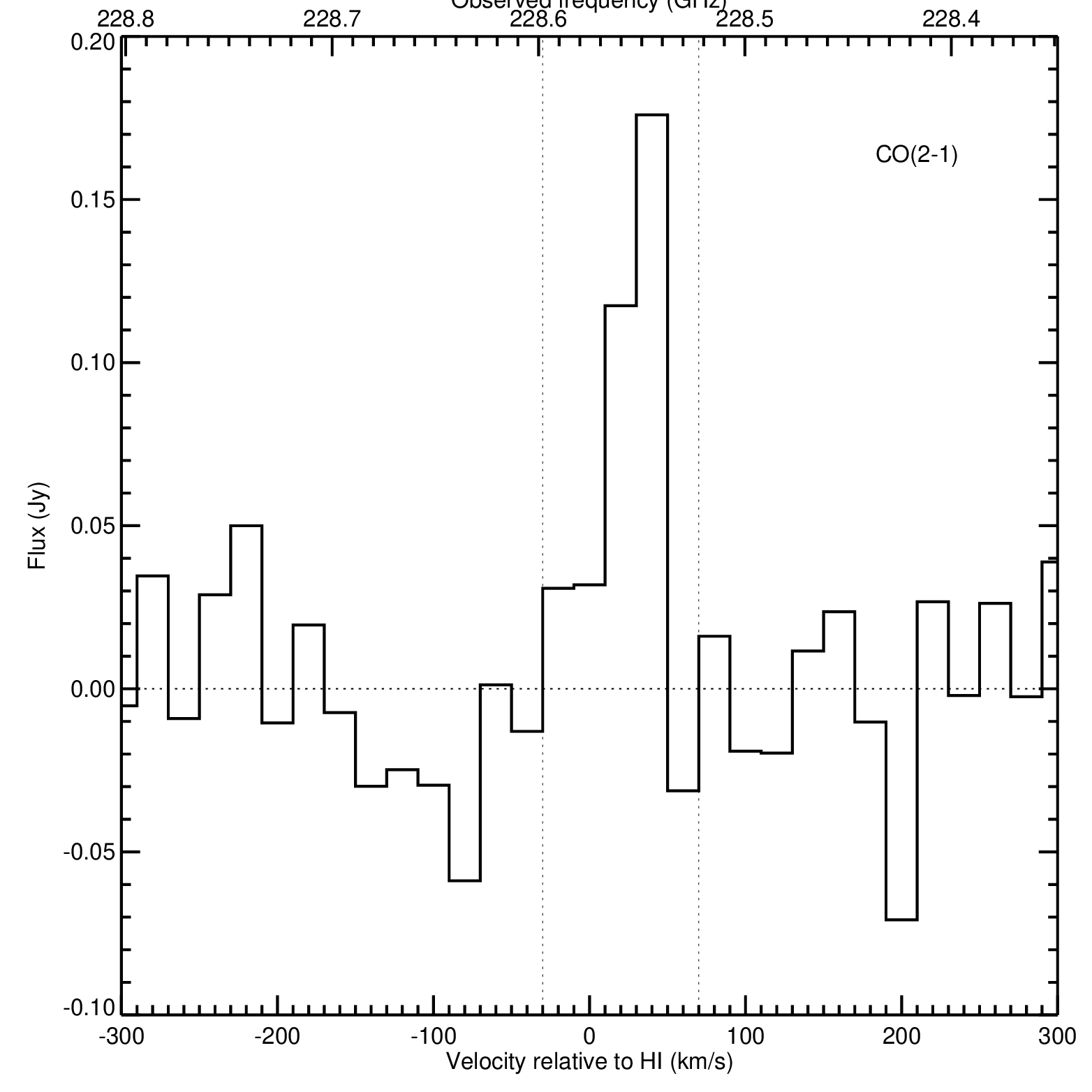}} &
\includegraphics[width=0.5\textwidth]{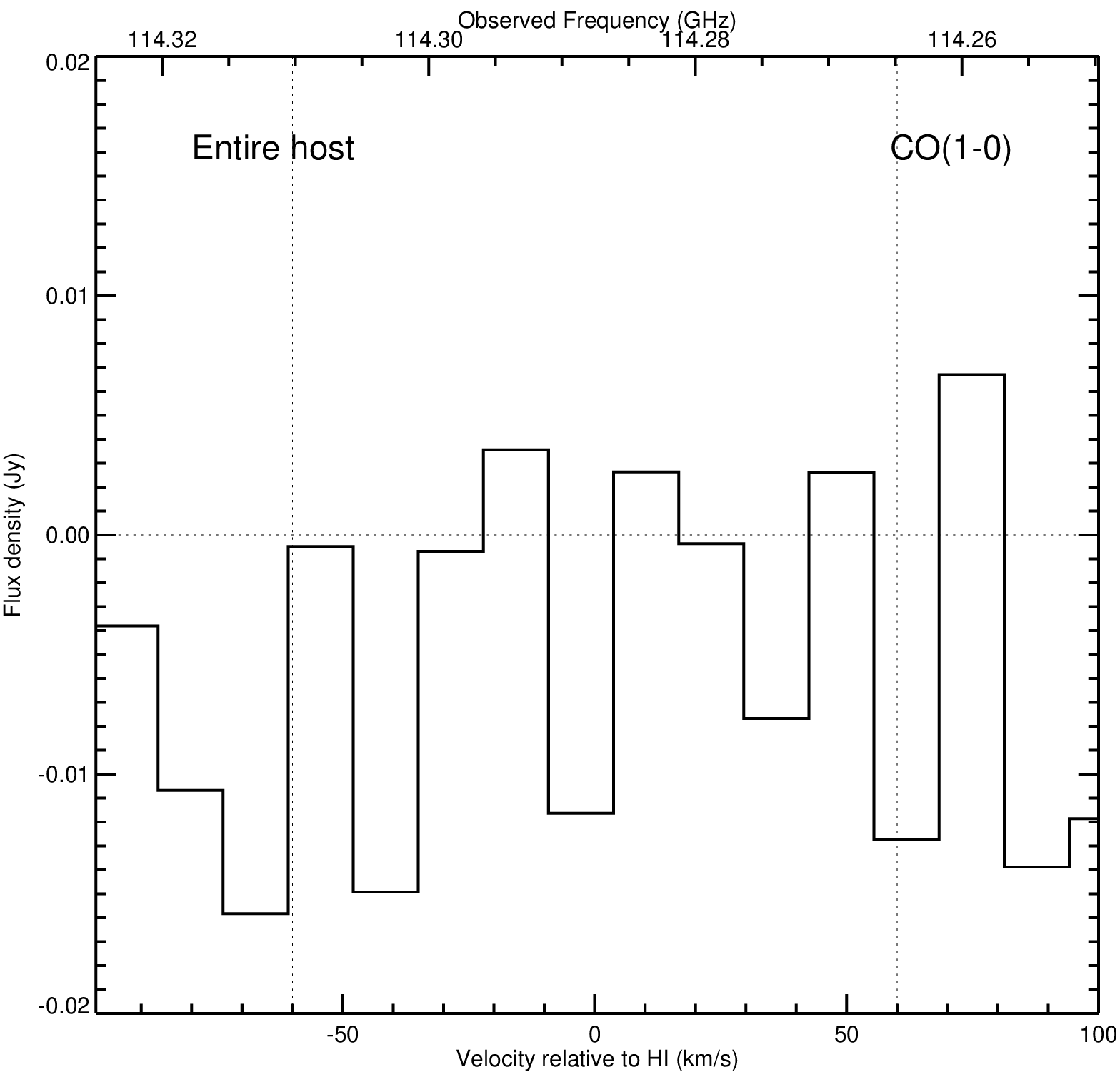} \\
\end{tabular}
\end{center}
\caption{{\cii} ({\it top left}), {\oi} ({\it top right}), CO(2-1) ({\it bottom left}), and CO(1-0) ({\it bottom right}) spectra of the entire host. The integrated fluxes were measured within inner dotted lines, and the errors within the outer  lines excluding the inner region.
}
\label{fig:hostspec}
\label{fig:co}
\end{figure*}

\begin{figure*}
\begin{center}
\includegraphics[width=0.9\textwidth]{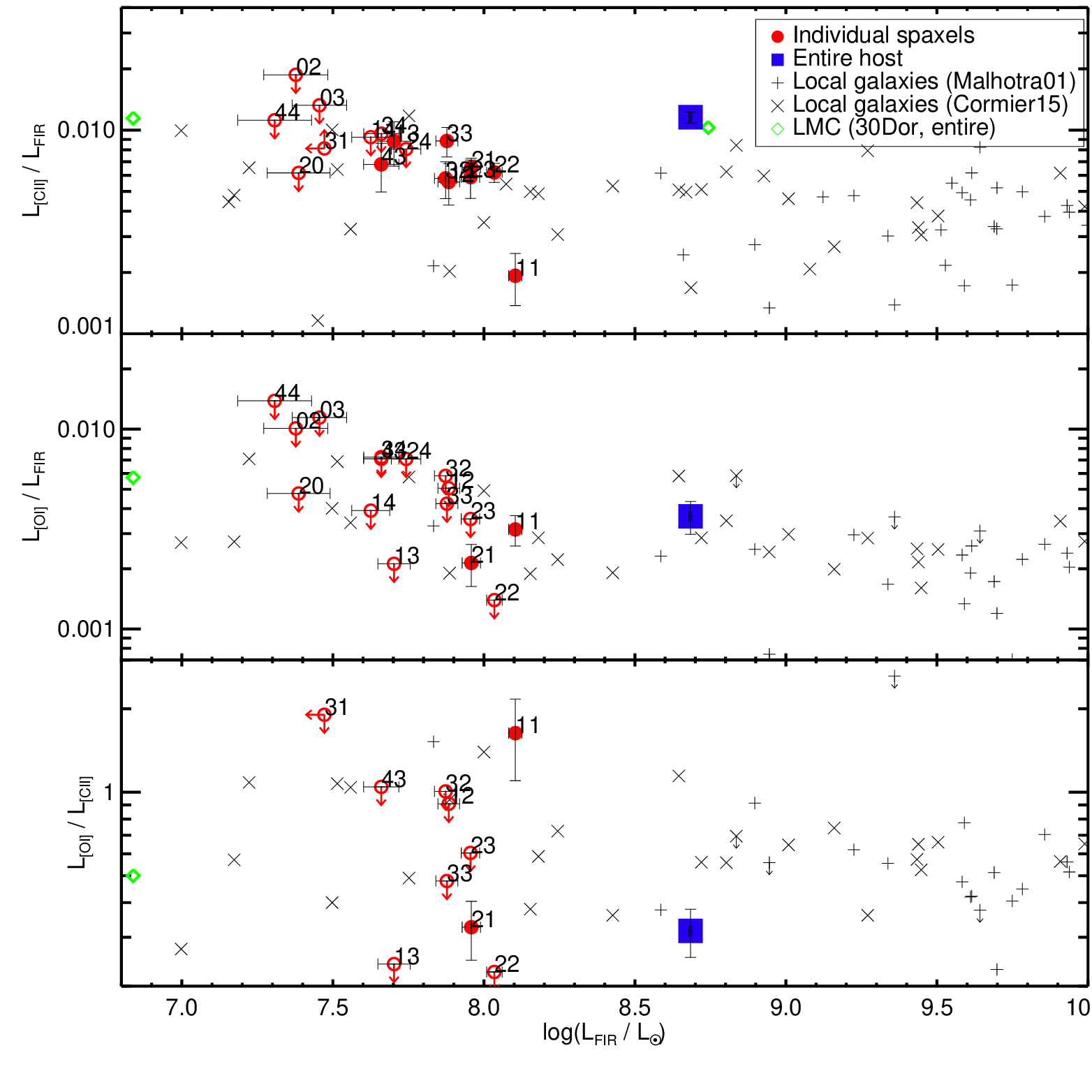} 
\end{center}
\caption{{\cii}-to-continuum ({\it top}), {\oi}-to-continuum ({\it middle}), and {\oi}-to-{\cii} ({\it bottom}) luminosity ratios  as a function of FIR luminosity ($40$--$120\,\micron$).  {\it Red circles} denote PACS spaxels, {\it blue square} denotes the entire host, whereas {\it plus signs}, {\it crosses} and {\it green diamonds} represent normal local galaxies \citep{malhotra01}, dwarf galaxies \citep{cormier15}, and the Large Magellanic Cloud \citep[the point at a lower luminosity is 30 Doradus and that at the higher luminosity is the entire galaxy;][]{poglitsch95,israel96}. 
The {\grb} host has elevated {\cii}/FIR and {\oi}/FIR ratios for its FIR luminosity.
}
\label{fig:lum}
\end{figure*}

\begin{figure}
\begin{center}
\includegraphics[width=0.5\textwidth,clip]{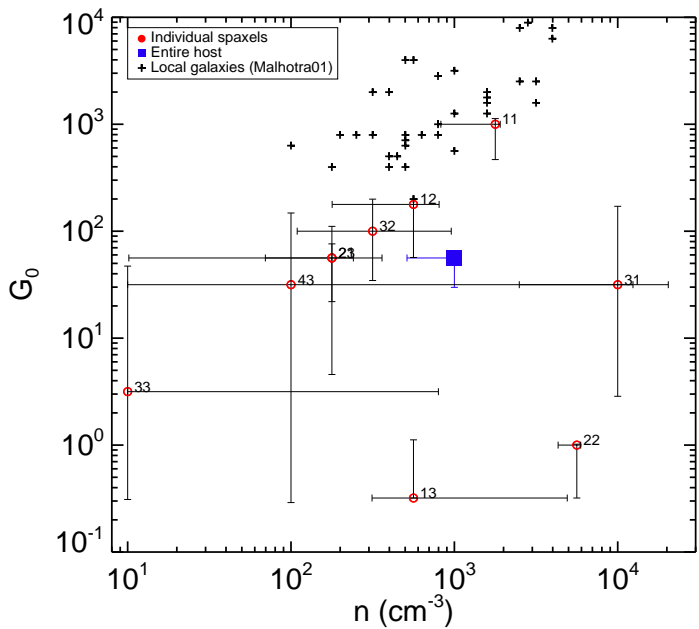} 
\end{center}
\caption{Gas density vs.~interstellar radiation field obtained from the PDR modelling.  {\it Red circles} denote PACS spaxels, {\it blue square} denotes the entire host, whereas {\it plus signs} represent normal local galaxies \citep{malhotra01}. 
The WR region (spaxel 11) exhibits high inferred radiation field and density.
}
\label{fig:ng}
\end{figure}

\begin{figure*}
\begin{center}
\includegraphics[width=0.9\textwidth]{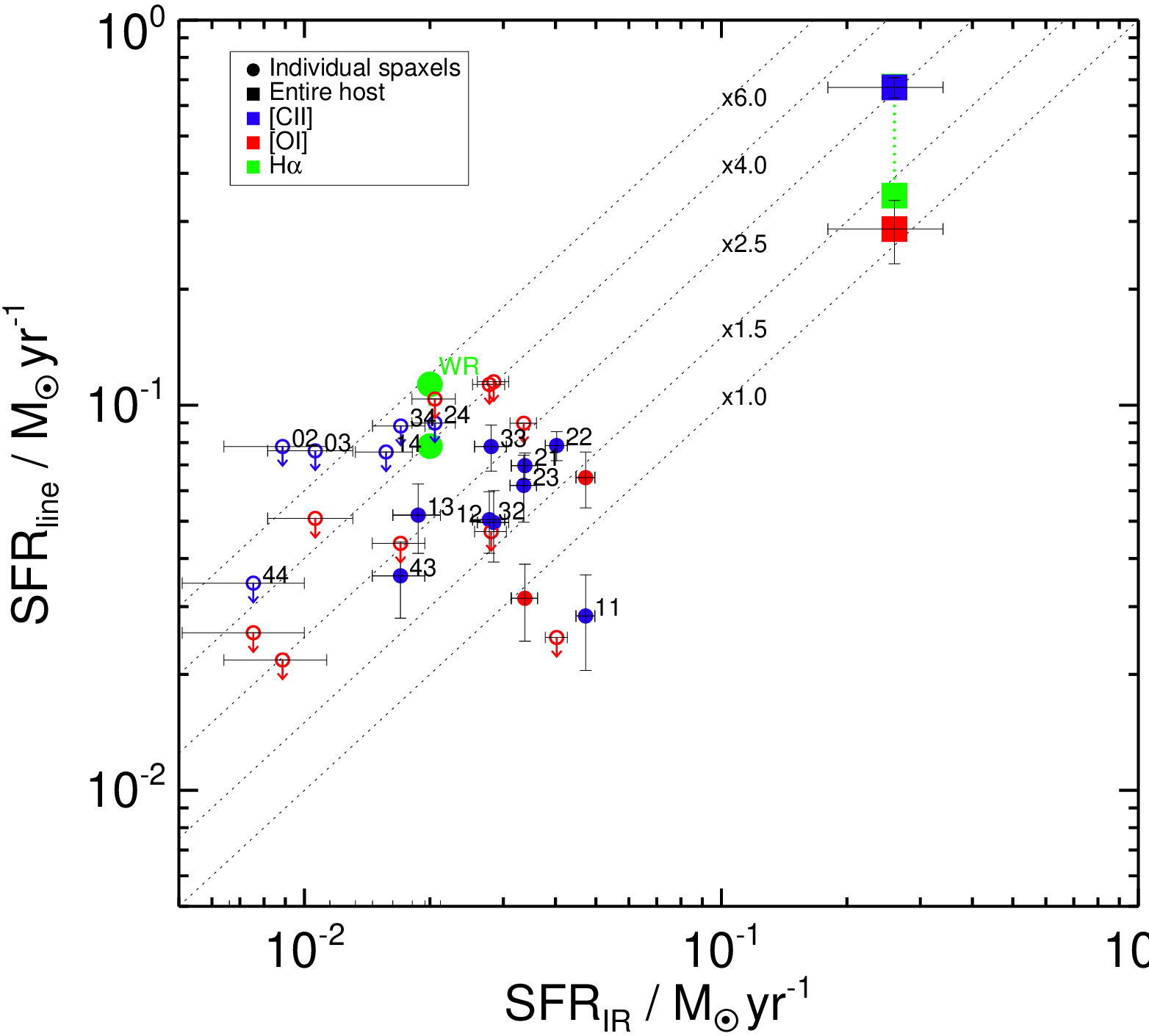} 
\end{center}
\caption{SFRs derived from IR luminosity ($8$-$1000\,\micron$; with the calibration of \citealt{kennicutt}) vs.~those derived from {\cii} ({\it blue}), {\oi} ({\it red}) (with the calibration of \citealt{delooze14}), and H$\alpha$ ({\it green}, higher values include dust correction as derived by \citealt{christensen08}) lines.  {\it Circles} denote PACS spaxels, and {\it squares} denote the entire host.  The  IR estimates are from \citet{michalowski14}. The H$\alpha$ estimate for the WR region is from \citet{christensen08}, and for the entire host from  \citet{sollerman05}.
{\it Dotted lines} show the locations at which SFR$_{\rm line}$ is higher than SFR$_{\rm IR}$ by the indicated factors.
The SFR derived from line ({\cii}, {\oi}, H$\alpha$) indicators are higher than from continuum (UV, IR, radio) indicators.
}
\label{fig:sfrvssfr}
\end{figure*}

\begin{figure}
\begin{center}
\includegraphics[width=0.5\textwidth]{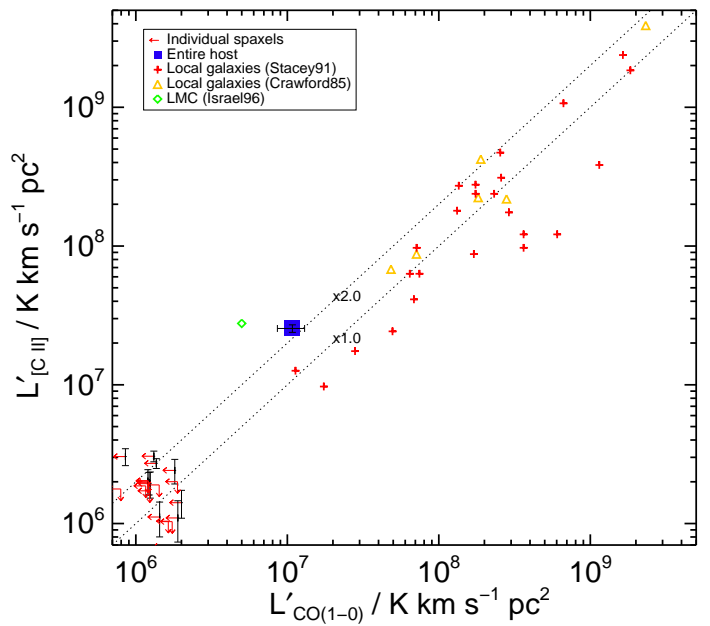} 
\end{center}
\caption{{\cii} luminosity as a function CO(1-0) luminosity of individual spaxels ({\it red arrows}), the entire {\grb} host ({\it blue square}), local galaxies \citep[{\it plus signs}, {\it diamonds}, and {\it crosses};][]{stacey91,crawford85} 
and the LMC \citep[{\it green diamond};][]{israel96}.
The {\grb} host has an elevated {\cii}/CO ratio compared with local galaxies.
}
\label{fig:ciico}
\end{figure}

\begin{table*}
\tiny
\centering
\caption{Physical properties of each spaxel and the entire host.\label{tab:phys}}
\medskip
\begin{tabular}{lrrrrrr}
\hline\hline
Reg & \multicolumn{1}{c}{SFR$_{\rm IR}$} & \multicolumn{1}{c}{SFR$_{\rm [CII]}$} & \multicolumn{1}{c}{SFR$_{\rm [OI]}$} & \multicolumn{1}{c}{$n$} & \multicolumn{1}{c}{$G_0$} & \multicolumn{1}{c}{$M_{\rm H_2, [CII]}$} \\
       & \multicolumn{1}{c}{(\msunyr)} & \multicolumn{1}{c}{(\msunyr)} & \multicolumn{1}{c}{(\msunyr)} & \multicolumn{1}{c}{(cm$^{-3}$)}     &      & \multicolumn{1}{c}{($10^6\msun$)} \\
(1)    & \multicolumn{1}{c}{(2)}     & \multicolumn{1}{c}{(3)}           & \multicolumn{1}{c}{(4)} & \multicolumn{1}{c}{(5)} & \multicolumn{1}{c}{(6)} & \multicolumn{1}{c}{(7)} \\
\hline
Host & $0.260\pm0.080$ &$ 0.668\pm0.040$ & $ 0.287\pm0.054$ & $1000_{-\phantom{11}487}^{+\phantom{11}34}$ &$56.2_{-\phantom{11}26.4}^{+\phantom{11}1.4}$ &$ 127.3\pm7.9$ \\
44 & $ 0.008\pm0.002$ & $ 0.012\pm0.007$ & $ 0.009\pm0.019$ & \multicolumn{1}{c}{$\cdots$}& \multicolumn{1}{c}{$\cdots$}& $ 2.3\pm1.4$ \\
43 & $ 0.017\pm0.002$ & $ 0.036\pm0.008$ & $ 0.015\pm0.019$ & $ 100_{-\phantom{1}\phantom{1}\phantom{1}90}^{+12302}$ & $ 31.6_{-\phantom{1}31.3}^{+116.3}$ & $ 7.1\pm1.6$ \\
42 & $ 0.006\pm0.002$ & $ 0.023\pm0.013$ & $ -0.012\pm0.012$ & \multicolumn{1}{c}{$\cdots$}& \multicolumn{1}{c}{$\cdots$}& $ 4.5\pm2.5$ \\
41 & $ -0.003\pm0.002$ & $ -0.034\pm0.020$ & $ -0.023\pm0.015$ & \multicolumn{1}{c}{$\cdots$}& \multicolumn{1}{c}{$\cdots$}& $ -6.8\pm3.9$ \\
40 & $ -0.000\pm0.002$ & $ -0.021\pm0.013$ & $ 0.007\pm0.022$ & \multicolumn{1}{c}{$\cdots$}& \multicolumn{1}{c}{$\cdots$}& $ -4.1\pm2.5$ \\
34 & $ 0.017\pm0.002$ & $ 0.029\pm0.011$ & $ -0.015\pm0.035$ & \multicolumn{1}{c}{$\cdots$}& \multicolumn{1}{c}{$\cdots$}& $ 5.8\pm2.1$ \\
33 & $ 0.028\pm0.002$ & $ 0.078\pm0.011$ & $ 0.016\pm0.018$ & $ 10_{-\phantom{1}\phantom{1}\phantom{1}\phantom{1}0}^{+\phantom{1}\phantom{1}789}$ & $ 3.2_{-\phantom{1}\phantom{1}2.9}^{+\phantom{1}44.0}$ & $ 15.2\pm2.1$ \\
32 & $ 0.028\pm0.002$ & $ 0.050\pm0.009$ & $ 0.038\pm0.016$ & $ 316_{-\phantom{1}\phantom{1}207}^{+\phantom{1}\phantom{1}641}$ & $ 100.0_{-\phantom{1}65.4}^{+\phantom{1}99.5}$ & $ 9.9\pm1.8$ \\
31 & $ 0.006\pm0.002$ & $ 0.028\pm0.009$ & $ 0.044\pm0.015$ & $ 10000_{-\phantom{1}7503}^{+10399}$ & $ 31.6_{-\phantom{1}28.7}^{+139.3}$ & $ 5.5\pm1.8$ \\
30 & $ 0.003\pm0.002$ & $ -0.009\pm0.011$ & $ 0.015\pm0.015$ & \multicolumn{1}{c}{$\cdots$}& \multicolumn{1}{c}{$\cdots$}& $ -1.7\pm2.2$ \\
24 & $ 0.021\pm0.002$ & $ 0.030\pm0.011$ & $ 0.035\pm0.014$ & \multicolumn{1}{c}{$\cdots$}& \multicolumn{1}{c}{$\cdots$}& $ 5.9\pm2.1$ \\
23 & $ 0.034\pm0.002$ & $ 0.062\pm0.012$ & $ 0.030\pm0.011$ & $ 178_{-\phantom{1}\phantom{1}168}^{+\phantom{1}\phantom{1}183}$ & $ 56.2_{-\phantom{1}51.6}^{+\phantom{1}54.8}$ & $ 12.1\pm2.4$ \\
22 & $ 0.040\pm0.002$ & $ 0.079\pm0.007$ & $ 0.008\pm0.008$ & $ 5620_{-\phantom{1}1302}^{+\phantom{1}\phantom{1}316}$ & $ 1.0_{-\phantom{1}\phantom{1}0.7}^{+\phantom{1}\phantom{1}0.0}$ & $ 15.3\pm1.4$ \\
21 & $ 0.034\pm0.002$ & $ 0.070\pm0.006$ & $ 0.032\pm0.007$ & $ 178_{-\phantom{1}\phantom{1}108}^{+\phantom{1}\phantom{1}\phantom{1}63}$ & $ 56.2_{-\phantom{1}34.3}^{+\phantom{1}19.9}$ & $ 13.6\pm1.1$ \\
20 & $ 0.009\pm0.002$ & $ -0.004\pm0.010$ & $ -0.000\pm0.010$ & \multicolumn{1}{c}{$\cdots$}& \multicolumn{1}{c}{$\cdots$}& $ -0.7\pm2.1$ \\
14 & $ 0.016\pm0.002$ & $ 0.025\pm0.010$ & $ -0.004\pm0.015$ & \multicolumn{1}{c}{$\cdots$}& \multicolumn{1}{c}{$\cdots$}& $ 5.0\pm2.0$ \\
13 & $ 0.019\pm0.002$ & $ 0.052\pm0.011$ & $ -0.011\pm0.014$ & $ 562_{-\phantom{1}\phantom{1}249}^{+\phantom{1}4347}$ & $ 0.3_{-\phantom{1}\phantom{1}0.0}^{+\phantom{1}\phantom{1}0.8}$ & $ 10.1\pm2.1$ \\
12 & $ 0.028\pm0.002$ & $ 0.050\pm0.010$ & $ 0.038\pm0.012$ & $ 562_{-\phantom{1}\phantom{1}383}^{+\phantom{1}\phantom{1}245}$ & $ 178.0_{-121.2}^{+\phantom{1}21.9}$ & $ 9.7\pm2.1$ \\
11 & $ 0.047\pm0.002$ & $ 0.028\pm0.008$ & $ 0.065\pm0.011$ & $ 1780_{-\phantom{1}\phantom{1}953}^{+\phantom{1}\phantom{1}129}$ & $ 1000.0_{-532.6}^{+129.9}$ & $ 5.6\pm1.6$ \\
10 & $ 0.005\pm0.002$ & $ -0.016\pm0.007$ & $ 0.009\pm0.018$ & \multicolumn{1}{c}{$\cdots$}& \multicolumn{1}{c}{$\cdots$}& $ -3.1\pm1.5$ \\
04 & $ 0.003\pm0.002$ & $ 0.020\pm0.014$ & $ -0.022\pm0.018$ & \multicolumn{1}{c}{$\cdots$}& \multicolumn{1}{c}{$\cdots$}& $ 3.9\pm2.7$ \\
03 & $ 0.011\pm0.002$ & $ 0.025\pm0.009$ & $ 0.017\pm0.018$ & \multicolumn{1}{c}{$\cdots$}& \multicolumn{1}{c}{$\cdots$}& $ 5.0\pm1.8$ \\
02 & $ 0.009\pm0.002$ & $ 0.026\pm0.013$ & $ 0.007\pm0.016$ & \multicolumn{1}{c}{$\cdots$}& \multicolumn{1}{c}{$\cdots$}& $ 5.1\pm2.5$ \\
01 & $ 0.004\pm0.002$ & $ 0.012\pm0.007$ & $ 0.017\pm0.014$ & \multicolumn{1}{c}{$\cdots$}& \multicolumn{1}{c}{$\cdots$}& $ 2.4\pm1.4$ \\
00 & $ 0.000\pm0.002$ & $ -0.016\pm0.013$ & $ -0.052\pm0.016$ & \multicolumn{1}{c}{$\cdots$}& \multicolumn{1}{c}{$\cdots$}& $ -3.2\pm2.6$ \\
\hline
\end{tabular}
\tablefoot{
(1) The entire host or the spaxel number. (2) Star formation rates (SFRs) from the total ($8$--$1000\,\micron$) luminosity using $\mbox{SFR}/\msunyr=10^{-10} L_{\rm IR}/\lsun$ assuming the \citet{chabrier03} initial mass function. (3), (4) SFRs from the {\cii} and {\oi} lines, respectively, using the calibration of \citet{delooze14}. (5) Gas density, (6) Strength of the interstellar radiation field in Habing unit (solar neighbourhood value), both derived via the PDR modelling (Sect.~\ref{sec:mod}). (7) Molecular gas mass estimated assuming $L'_{\rm CO(1-0)}= L'_{\cii}$ (see Sect.~\ref{sec:mod}) and the Galactic CO-to-{\htwo} conversion factor $\alpha_{\rm CO}=5\msun/(\Kkmspc)$.
}
\end{table*}

\begin{table*}
\centering
\caption{APEX CO(2-1) line fluxes and luminosities.\label{tab:co}}
\medskip
\begin{tabular}{lcccccc}
\hline\hline
Reg & \multicolumn{1}{c}{$F_{\rm int}$} & S/N & \multicolumn{1}{c}{$F_{\rm int}$} & \multicolumn{1}{c}{$L$} & \multicolumn{1}{c}{$\lp$} & \multicolumn{1}{c}{$M_{\rm H_2, CO}$}   \\
       & \multicolumn{1}{c}{(Jy km s$^{-1}$)} &  & \multicolumn{1}{c}{($10^{-17}$W m$^{-2}$)}    & \multicolumn{1}{c}{($10^{5}L_\odot$)} & \multicolumn{1}{c}{($10^{6}\mbox{K km s}^{-1} \mbox{ pc}^2$)} &\multicolumn{1}{c}{($10^6\msun$)}   \\
(1)    & \multicolumn{1}{c}{(2)}     & (3)  & \multicolumn{1}{c}{(4)}           & \multicolumn{1}{c}{(5)} & \multicolumn{1}{c}{(6)} & \multicolumn{1}{c}{(7)} \\
\hline
Host & $ 6.5\pm1.3$ & $ 5.1$ & $ 0.0050\pm0.0010$ & $ 0.021\pm0.004$ & $ 5.4\pm1.1$ & $ 54.2\pm10.7$ \\ 
\hline
\end{tabular}
\tablefoot{
(1) The entire host. (2) Integrated flux within dotted lines on Fig.~\ref{fig:co}. (3) Signal-to-noise ratio of the line. (4) Integrated flux in W m$^{-2}$. (5) line luminosity in solar luminosity. (6) line luminosity using equation 3 in \citet{solomon97}. (7) Molecular gas mass estimated assuming $L'_{\rm CO(1-0)}= 2\times L'_{\rm CO(2-1)}$ (see Sect.~\ref{sec:mod}) and the Galactic CO-to-{\htwo} conversion factor $\alpha_{\rm CO}=5\msun/(\Kkmspc)$.
}
\end{table*}

\begin{table*}
\centering
\caption{ALMA CO(1-0) line fluxes and luminosities of each spaxel and the WR region.\label{tab:fluxCO}}
\medskip
\begin{tabular}{lrrrrrrr}
\hline\hline
Reg & \multicolumn{1}{c}{$F_{\rm int}$} & S/N & \multicolumn{1}{c}{$F_{\rm int}$} & \multicolumn{1}{c}{$L$} & \multicolumn{1}{c}{$\lp$} & \multicolumn{1}{c}{$M_{\htwo\, {\rm CO}}$}  \\
       & \multicolumn{1}{c}{(Jy km s$^{-1}$)} &  & \multicolumn{1}{c}{($10^{-22}$W m$^{-2}$)}    & \multicolumn{1}{c}{($L_\odot$)} & \multicolumn{1}{c}{($10^{6}\mbox{K km s}^{-1} \mbox{ pc}^2$)} & \multicolumn{1}{c}{($10^6\msun$)}  \\
(1)    & \multicolumn{1}{c}{(2)}     & (3)  & \multicolumn{1}{c}{(4)}           & \multicolumn{1}{c}{(5)} & \multicolumn{1}{c}{(6)} & \multicolumn{1}{c}{(7)}  \\
\hline
Host & $ -0.406\pm0.316$ & $ -1.3$ & $ -15.6\pm12.1$ & $ -66.4\pm51.7$ & $ -1.36\pm1.05$ & $ -6.8\pm5.3$ \\
WR & $ 0.058\pm0.049$ & $ 1.2$ & $ 2.2\pm1.9$ & $ 9.5\pm\phantom{1}8.1$ & $ 0.19\pm0.16$ & $ 1.0\pm0.8$ \\
44 & $ 0.271\pm0.125$ & $ 2.2$ & $ 10.4\pm4.8$ & $ 44.4\pm20.4$ & $ 0.91\pm0.42$ & $ 4.5\pm2.1$ \\
43 & $ 0.318\pm0.141$ & $ 2.3$ & $ 12.2\pm5.4$ & $ 52.1\pm23.0$ & $ 1.06\pm0.47$ & $ 5.3\pm2.3$ \\
42 & $ 0.209\pm0.109$ & $ 1.9$ & $ 8.1\pm4.2$ & $ 34.3\pm17.8$ & $ 0.70\pm0.36$ & $ 3.5\pm1.8$ \\
41 & $ -0.040\pm0.132$ & $ -0.3$ & $ -1.5\pm5.1$ & $ -6.5\pm21.6$ & $ -0.13\pm0.44$ & $ -0.7\pm2.2$ \\
40 & $ 0.099\pm0.128$ & $ 0.8$ & $ 3.8\pm4.9$ & $ 16.3\pm21.0$ & $ 0.33\pm0.43$ & $ 1.7\pm2.1$ \\
34 & $ 0.301\pm0.132$ & $ 2.3$ & $ 11.6\pm5.1$ & $ 49.2\pm21.6$ & $ 1.00\pm0.44$ & $ 5.0\pm2.2$ \\
33 & $ 0.003\pm0.126$ & $ 0.0$ & $ 0.1\pm4.9$ & $ 0.5\pm20.7$ & $ 0.01\pm0.42$ & $ 0.1\pm2.1$ \\
32 & $ 0.118\pm0.124$ & $ 0.9$ & $ 4.5\pm4.8$ & $ 19.3\pm20.4$ & $ 0.39\pm0.42$ & $ 2.0\pm2.1$ \\
31 & $ 0.288\pm0.140$ & $ 2.1$ & $ 11.1\pm5.4$ & $ 47.1\pm22.8$ & $ 0.96\pm0.47$ & $ 4.8\pm2.3$ \\
30 & $ 0.263\pm0.111$ & $ 2.4$ & $ 10.1\pm4.3$ & $ 43.1\pm18.2$ & $ 0.88\pm0.37$ & $ 4.4\pm1.9$ \\
24 & $ 0.154\pm0.103$ & $ 1.5$ & $ 5.9\pm4.0$ & $ 25.1\pm16.9$ & $ 0.51\pm0.34$ & $ 2.6\pm1.7$ \\
23 & $ 0.267\pm0.137$ & $ 2.0$ & $ 10.3\pm5.3$ & $ 43.7\pm22.4$ & $ 0.89\pm0.46$ & $ 4.5\pm2.3$ \\
22 & $ 0.178\pm0.108$ & $ 1.6$ & $ 6.8\pm4.2$ & $ 29.1\pm17.7$ & $ 0.59\pm0.36$ & $ 3.0\pm1.8$ \\
21 & $ 0.170\pm0.119$ & $ 1.4$ & $ 6.6\pm4.6$ & $ 27.9\pm19.5$ & $ 0.57\pm0.40$ & $ 2.8\pm2.0$ \\
20 & $ 0.225\pm0.121$ & $ 1.9$ & $ 8.7\pm4.7$ & $ 36.9\pm19.9$ & $ 0.75\pm0.41$ & $ 3.8\pm2.0$ \\
14 & $ 0.009\pm0.114$ & $ 0.1$ & $ 0.4\pm4.4$ & $ 1.5\pm18.7$ & $ 0.03\pm0.38$ & $ 0.2\pm1.9$ \\
13 & $ 0.102\pm0.129$ & $ 0.8$ & $ 3.9\pm5.0$ & $ 16.7\pm21.1$ & $ 0.34\pm0.43$ & $ 1.7\pm2.2$ \\
12 & $ 0.156\pm0.109$ & $ 1.4$ & $ 6.0\pm4.2$ & $ 25.5\pm17.8$ & $ 0.52\pm0.36$ & $ 2.6\pm1.8$ \\
11 & $ 0.208\pm0.111$ & $ 1.9$ & $ 8.0\pm4.3$ & $ 34.1\pm18.1$ & $ 0.70\pm0.37$ & $ 3.5\pm1.9$ \\
10 & $ 0.116\pm0.114$ & $ 1.0$ & $ 4.4\pm4.4$ & $ 18.9\pm18.6$ & $ 0.39\pm0.38$ & $ 1.9\pm1.9$ \\
04 & $ 0.114\pm0.117$ & $ 1.0$ & $ 4.4\pm4.5$ & $ 18.6\pm19.2$ & $ 0.38\pm0.39$ & $ 1.9\pm2.0$ \\
03 & $ 0.124\pm0.123$ & $ 1.0$ & $ 4.8\pm4.7$ & $ 20.3\pm20.1$ & $ 0.41\pm0.41$ & $ 2.1\pm2.1$ \\
02 & $ 0.130\pm0.116$ & $ 1.1$ & $ 5.0\pm4.5$ & $ 21.3\pm19.0$ & $ 0.43\pm0.39$ & $ 2.2\pm1.9$ \\
01 & $ 0.248\pm0.121$ & $ 2.0$ & $ 9.5\pm4.7$ & $ 40.5\pm19.9$ & $ 0.83\pm0.41$ & $ 4.1\pm2.0$ \\
00 & $ 0.160\pm0.115$ & $ 1.4$ & $ 6.2\pm4.4$ & $ 26.2\pm18.9$ & $ 0.53\pm0.39$ & $ 2.7\pm1.9$ \\
\hline
\end{tabular}
\tablefoot{
(1) The entire host, the WR region, or the spaxel number. (2) Integrated flux within dotted lines on Fig.~\ref{fig:cospax} and \ref{fig:coWR}. (3) Signal-to-noise ratio of the line. (4) Integrated flux in W m$^{-2}$. (5) Line luminosity in solar luminosity. (6) Line luminosity using equation 3 in \citet{solomon97}. (7) Molecular gas mass estimated assuming the Galactic CO-to-{\htwo} conversion factor $\alpha_{\rm CO}=5\msun/(\Kkmspc)$.
}
\end{table*}

The resulting {\cii} and {\oi} spectra in each PACS spaxel are presented in the top row of Fig.~\ref{fig:spax}, whereas the summed spectra of the entire galaxy are shown in Fig.~\ref{fig:hostspec}.
The integrated fluxes were measured within inner dotted lines, and the errors within the outer lines excluding the inner region. Fluxes and luminosities are presented in Table~\ref{tab:flux}.
The flux spatial distributions are shown in the middle row of Fig.~\ref{fig:spax}. The {\cii} flux distribution follows a normal radial dependence with the brightest spaxel at the center of the galaxy. In contrast, most of the {\oi} flux is not at the galaxy centre, but is concentrated close to the WR region (spaxel 11). 

The {\cii} and {\oi} line and line-to-continuum ratios of all spaxels and the entire host are shown in Fig.~\ref{fig:lum}, and compared with local normal star-forming galaxies \citep{malhotra01}, dwarf galaxies \citep{cormier15} from the  Dwarf Galaxy Survey \citep{madden13} and the Large Magellanic Cloud \citep[LMC;][]{poglitsch95,israel96}. The {\grb} host has elevated {\cii}/FIR and {\oi}/FIR ratios for its FIR luminosity and a low {\oi}/{\cii} ratio. 

The results of the PDR modelling are shown in the bottom row of Fig.~\ref{fig:spax} and in Fig.~\ref{fig:ng}.  The derived densities and radiation field strengths together with SFRs and molecular gas masses are presented in Table~\ref{tab:phys}. Spaxel 11 has the highest {\oi}/{\cii} luminosity ratio ($\sim1.7$)  among all regions in the host, which leads to the highest radiation field of $G_0\sim1000$. Spaxels near the WR regions in the southwestern part of the galaxy (number 11, 22, 31) have also the highest densities in excess of $1000\,\mbox{cm}^{-3}$.

Fig.~\ref{fig:sfrvssfr} shows the SFRs derived using luminosities of emission lines:  {\cii}, {\oi} \citep[the calibration of ][]{delooze14} and H$\alpha$ \citep[as reported by][]{sollerman05,christensen08} as a function of SFRs derived from total IR ($8$--$1000\,\micron$) continuum emission. All line estimates are $\sim1.5$--$6$ times larger than the IR estimates (and also than the UV estimate; see \citealt{michalowski14}). This is robust, because we are comparing the line estimates to the relatively high value of $0.26\,\msunyr$ obtained by SED modelling in \citet{michalowski14}. If we used the calibration of \citet{kennicutt}, then we would obtain $0.1\,\msunyr$, whereas that of \citet[][their eq.~18]{murphy11} would give $0.17\,\msunyr$.  On the other hand, if, instead of the calibration of \citet{delooze14}, we used that of \citet{sargsyan12}, then for the entire host we would obtain SFR$_{\cii}\sim0.465\pm0.023\,\msunyr$, also a factor of $\sim1.7$ higher than the IR estimate. In contrast, the calibration of \citet{herreracamus15} gives SFR$_{\cii}\sim0.115\pm0.005\,\msunyr$, lower than SFR$_{\rm IR}$. It is unclear why this calibration is a factor of $\sim6$ lower than that of \citet{delooze14}. Yet, we will use the latter given the larger sample size it is based on (530 vs.~46), and because the {\cii}/FIR luminosity ratio of the {\grb} host is indeed  larger than that of other galaxies (Fig.~\ref{fig:lum}), so it is expected that $\mbox{SFR}_{\cii}>\mbox{SFR}_{\rm IR}$.

The APEX CO(2-1) spectrum of the entire host is shown in Fig.~\ref{fig:co} and the CO flux, luminosity and the resulting molecular gas mass are presented in Table~\ref{tab:co}.
The CO-derived molecular mass of $\sim5.4\times10^7\,\msun$ is lower than the {\cii}-derived molecular mass of $\sim1.3\times10^8\,\msun$ (both estimates are based on a similar area over which the emission is summed, see Fig.~\ref{fig:thumb}).Both estimates are lower than the upper limit of $3\times10^8\,\msun$ derived by \citet{hatsukade07}.  Fig.~\ref{fig:ciico} demonstrates the high {\cii}/CO(1-0) luminosity ratio of the {\grb} host compared with local galaxies \citep{crawford85,stacey91}. 

The ALMA CO(1-0) spectra extracted from the extents of PACS spaxels are shown in Fig.~\ref{fig:cospax}, whereas the spectrum of the WR region extracted within a 1.7\arcsec\ radius aperture is shown in Fig.~\ref{fig:coWR}. No significant emission is detected, so we measured fluxes within $40\,\kms$ width (the {\hi} velocity width at a given position; fig.~5 of \citealt{arabsalmani15b}) at the frequency giving the highest conservative upper limit. These frequencies are shown as dotted lines on  Figs.~\ref{fig:cospax}, and \ref{fig:coWR}. 
Fig.~\ref{fig:co} shows the CO(1-0) spectrum of the entire host extracted within the region marked in Fig.~\ref{fig:thumb} as a green circle. However, this region has a diameter of $\sim27\arcsec$, whereas our ALMA observations are not sensitive to smooth emission extended over scales of more than $\sim11\arcsec$. Hence, if such emission is present, then our CO(1-0) flux of the entire host is underestimated. Indeed the integrated flux ratio of $S_{\rm CO(2-1)}/S_{\rm CO(1-0)}>10$ ($2\sigma$) is indeed higher than that of any other galaxy \citep{carilli13}, so our ALMA observations do not seem to probe the main emission component in the {\grb} host.
The CO(1-0) fluxes, luminosities, and the resulting molecular gas masses are listed in Table~\ref{tab:fluxCO}. Again, Fig.~\ref{fig:ciico} shows high {\cii}/CO luminosity ratio of the spaxels which are detected at {\cii}.

The resulting star formation efficiency  (SFE) is  $L_{\rm IR}/L'_{\rm CO(1-0)}=(95\pm19)\,\lsun/(\Kkmspc)$ (assuming $L'_{\rm CO(1-0)}= 2\times L'_{\rm CO(2-1)}$). 
This is higher than the average for local spirals of $\sim(48\pm7)\,\lsun/(\Kkmspc)$ derived by \citet[][their fig.~13]{daddi10}, close to the top-end of the distribution. 
Together with three GRB hosts reported to be CO-deficient \citep{hatsukade14,stanway15}, the picture emerges that GRB hosts exhibit the lowest CO luminosities among star-forming galaxies, given their SFRs.
On the other hand, if we assume that {\cii} is a good tracer of molecular gas, then $\mbox{SFE}=L_{\rm IR}/L'_{\cii}=(40.4\pm2.0)\,\lsun/(\Kkmspc)$, in agreement with local galaxies.

 In order to quantify further the potential molecular-gas deficiency of the {\grb} host, we used the relation between the metallicity, atomic gas and molecular gas for dwarf galaxies  provided by \citet[][their section 4]{filho16}, based on the calibration of \citet{amorin16}: $\log({\mhtwo}) = 1.2 \log({\mhi}) - 1.5\times [12 + \log(\mbox{O}/\mbox{H}) - 8.7] - 2.2$. For its atomic gas mass $\log(\mhi/\msun)\sim8.849$ \citep{michalowski15hi} and metallicity $12 + \log(\mbox{O}/\mbox{H})\sim8.6$ \citep{sollerman05}, the {\grb} host should have $\mhtwo\sim4\times10^8\,\msun$, $\sim7$ times higher than the CO estimate (Table~\ref{tab:co}) and $\sim3$ times higher than the {\cii} estimate (Table~\ref{tab:phys}). Hence, the {\grb} host has a low molecular gas mass for its atomic gas mass and metallicity. 
  A similar conclusion can be obtained from the relation between SFR, CO luminosity and metallicity presented in \citet[][their fig.~5]{hunt15}: $\log(\mbox{SFR}/\lp_{\rm CO}) =-2.25\times[12+\log(\mbox{O}/\mbox{H})]+11.31$. According to this relation $\mbox{SFR}/\lp_{\rm CO}$ of the {\grb} host should be equal to $9.1\times10^{-9}\,\msunyr / (\Kkmspc)$, whereas using SFR$_{\rm IR}$ the measured value is $\sim2.5$ times higher: $24\pm9\times10^{-9}\,\msunyr / (\Kkmspc)$, indicating low CO luminosity for its SFR and metallicity.

\section{Discussion}
\label{sec:discussion}

\begin{figure*}
\begin{center}
\includegraphics[width=\textwidth]{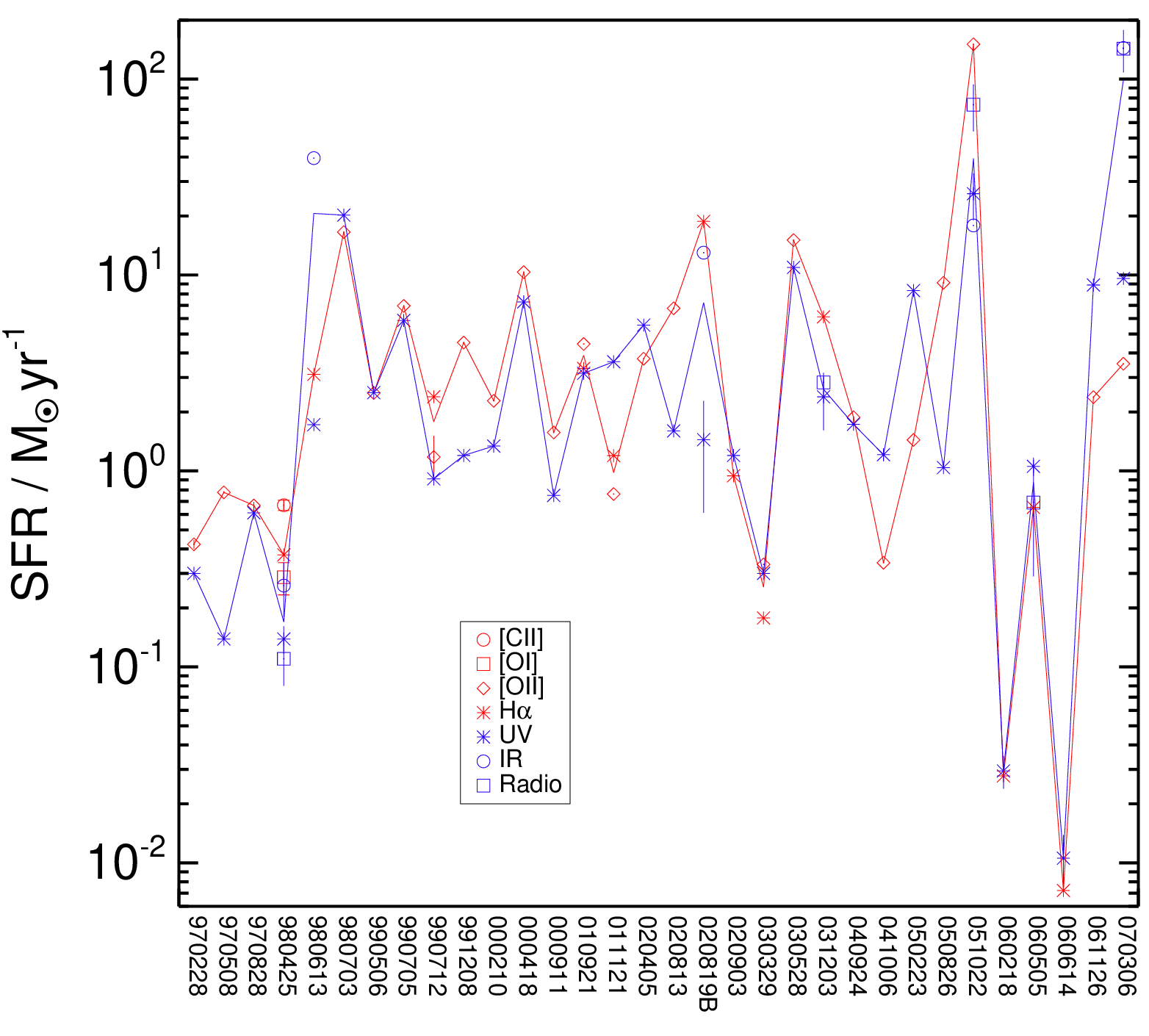} 
\end{center}
\caption{The compilation of SFRs measured from emission lines ({\it red}; {\it circles}: {\cii}, {\it squares}: {\oi}, {\it diamonds}: {\oii},  {\it asterisks}: H$\alpha$) and continuum indicators ({\it blue}; {\it asterisks}: ultraviolet,  {\it circles}: infrared, {\it squares}: radio). The x-axis position corresponds to different GRB hosts. The {\it red} and {\it blue lines} show the average line- and continuum-based SFR for a given GRB host, respectively.
The line-based indicators give on average systematically higher estimates.
}
\label{fig:sfrall}
\end{figure*}

\subsection{Recent inflow of atomic gas from the intergalactic medium triggering star formation}

Summarising, we have the following pieces of information about the {\grb} host:
{\it i)} it has elevated {\cii}/FIR and {\oi}/FIR ratios (Fig.~\ref{fig:lum}) and higher values of SFRs derived from line ({\cii}, {\oi}, H$\alpha$) than from continuum (UV, IR, radio) indicators (Fig.~\ref{fig:sfrvssfr});
{\it ii)} its {\cii} emission exhibits a normal radial profile, whereas  {\oi} emission is concentrated close to the WR region (Fig.~\ref{fig:spax}), leading to high inferred radiation field and density (Fig.~\ref{fig:ng}).
{\it iii)} its CO luminosity is at the lower end of the distribution for other galaxies, leading to an elevated {\cii}/CO ratio (Fig.~\ref{fig:ciico}) and {\mhtwo} derived from CO lower by a factor of $\sim2$ than that inferred from {\cii}.

All these observables can be explained by the hypothesis presented in \citet{michalowski15hi} that this galaxy in particular, and possibly most GRB hosts, experienced a  very recent inflow of atomic gas triggering star formation. Indeed, {\hi} is concentrated close to the position of the WR region \citep[our spaxels 11, 12, 21;][]{arabsalmani15b}. In this scenario the newly-acquired atomic gas quickly becomes cool and dense, leading to intense star formation (giving rise to the birth of the {\grb} progenitor), and explaining the high {\oi} and high derived radiation field and density. The accretion event disturbs the gas reservoir in other parts of the galaxy as well, so the total SFR and ionised carbon emission are enhanced. However, dust has not had time to build up and the reprocessing of stellar emission by dust is much slower (see below), so the infrared emission is still low giving rise to high {\cii}/FIR and $\mbox{SFR}_{\cii}/\mbox{SFR}_{\rm IR}$ ratios. 

We note that high gas densities close to the GRB position (Fig.~\ref{fig:ng}) are consistent with the conclusion of \citet{michalowski14} based on the properties of the WR region. FIR spectroscopy of other GRB hosts is needed to test whether high density and radiation field is a condition necessary for GRB explosions.

Our {\cii} and {\oi} maps are not of enough resolution to investigate the spatial distribution of the  resulting specific SFR ($\mbox{sSFR}\equiv\mbox{SFR}/M_*$). However, stellar mass is distributed relatively smoothly across the host, and in particular the WR region does not have an exceptionally high or low mass \citep{christensen08,michalowski09}. Hence, SFR and sSFR distributions based on H$\alpha$ emission are very similar \citep{christensen08}.

Our data allows us to estimate the timescale of the required gas inflow.
The line emission ({\cii}, {\oi}, H$\alpha$) is connected with the most massive (O type) stars, so it traces very recent star formation, happening during the last $\sim10$\,Myr. On the other hand, UV, IR and radio emission traces the average SFR over the last  $\sim100$\,Myr \citep{kennicutt}, so the continuum emission is relatively insensitive to the enhancement in star formation due to very recent gas inflow. Hence, our hypothesis requires the inflow to happen at most a few tens of Myr ago, in order that the line SFR indicators still give higher values than the continuum ones.

The timescale of less than a few tens of Myr is consistent with the estimates of the age of the GRB progenitor for {\grb} in particular, and for all GRBs in general. \citet{sollerman05} obtained the stellar age of the region in which {\grb} exploded of $\sim6$\,Myr,  
which is consistent with the timescale we propose for the trigger of star formation.  Similarly, if the progenitor of {\grb} was a runaway star expelled from the WR region \citep{hammer06,vandenheuvel13}, then it would need only $\sim3$--$6$\,Myr to reach its explosion site assuming a reasonable kick velocity of $\sim130$--$260\,\kms$. Similar ages of stellar progenitors were derived for other GRBs \citep{thone08,ostlin08}.

On the other hand, the free-fall time is 
\begin{equation}
t_{\rm ff}=\left(\frac{3\pi}{32G\rho}\right)^{1/2}  \approx1.6\,\mbox{Myr}\left(\frac{n}{1000\,\mbox{cm}^{-3}}\right)^{-1/2}
\end{equation}
where $G$ is the gravitational constant, $\rho$ is gas density, and $n$ is the number density for which we assumed $\rho=n m_p$, where $m_p$ is the proton mass (i.e.~assuming atomic gas only).
For our measured density of $n\sim1000\,\mbox{cm}^{-3}$ (Fig.~\ref{fig:ng} and Table~\ref{tab:phys}) the free-fall time is a few Myr. Star formation starts after one to a few $t_{\rm ff}$, so if, as proposed above, the inflow happened a few tens of Myr ago, then this is long enough for star formation to  start.

We investigate the relation between SFRs of other GRB hosts derived from line and continuum emission  in Fig.~\ref{fig:sfrall}, in which we compiled the H$\alpha$, {\oii}$\lambda3727$, UV, IR and radio estimates \citep{bloom,bloom01,djorgovski01,djorgovski03,price02,garnavich03,christensen04, prochaska04,gorosabel05,sollerman05,sollerman06,castroceron06,castroceron10,dellavalle06,thone08,michalowski09,michalowski12grb,michalowski14,michalowski15hi, savaglio09,levesque10b,levesque10c,stanway10,watson11,hjorth12,jakobsson12,perley13b,perley15,hunt14,schady14}, and converted them to the \citet{chabrier03} IMF if needed (we divided by 1.8 the estimates based on the \citealt{salpeter} IMF).
We note that the {\oii}$\lambda3727$ estimator is strongly metallicity-dependent \citep{kewley04}, but in our sample these estimates are consistent with those from H$\alpha$ (mean ratio of $1.1\pm0.32$).

In principle it would also be advantageous to investigate sSFRs of GRB hosts. This would not change line-to-continuum ratios, but, if done properly, it could decrease the scatter on Fig.~\ref{fig:sfrall}. This is because stellar mass estimates depend on many assumptions like star formation histories, initial mass function, stellar models \citep{michalowski12mass,michalowski14mass,pacifici12,mitchell13,simha14}, which may not be universal in the sample of GRB hosts. However,  the inhomogeneity of optical and near-infrared data for this sample renders this investigation beyond the scope of this paper.

In 20 out of 32 cases the line SFR is higher than the continuum SFR. The mean ratio of the SFR derived from line indicators to that from the continuum indicator is $1.74\pm0.32$. 
This is consistent with a very recent enhancement of star formation (reflected in stronger line emission, but not influencing the continuum yet) being common
among GRB hosts. Hence, galaxies which have recently experienced  inflow of gas may preferentially host stars exploding as GRBs. This may be due to the GRB metallicity bias favouring metal-poor environments \citep{yoon05,yoon06,woosley06,piranomonte15,schulze15,trenti15,vergani15,japelj16,perley16b}. Hence the accretion of the metal-poor gas from the intergalactic medium is likely required to produce regions with low enough metallicity.
Therefore GRBs may be used to select unique samples of galaxies suitable for the investigation of recent gas accretion.

The line and continuum indicators have been calibrated so that they are consistent with each other on average for star-forming galaxies \citep[see e.g.][]{wijesinghe11,davies16, wang16}, so the higher line-based SFRs of GRB hosts are not due to systematic effects. Our interpretation of strong line emission is consistent with the investigation of other  galaxies.
The influence of the star formation history on the measured SFRs with various indicators was investigated by \citet{guo16}, who interpreted low far-UV-to-H$\alpha$ SFR ratio of star-forming galaxies as a sign of recent starburst, because this ratio anti-correlates with specific SFR measured with H$\alpha$ (their fig.~6, see also \citealt{sullivan00,iglesias04,boselli09,lee09b,meurer09,fumagalli11,weisz12,dasilva14}).

\citet{michalowski15hi} proposed that star formation may proceed directly in the accreted atomic gas before the conversion to molecular gas. This is possible, because the cooling timescale is much shorter than the {\hi}-to-{\htwo} conversion timescale, so just after the atomic gas is accreted it can start forming stars before it converts to the molecular phase  \citep{glover12,krumholz12,hu16}.  
As a result, molecular gas mass of GRB hosts is lower than what would be expected from their SFRs (which is partially fuelled by atomic gas). 

\subsection{Alternative explanations}

Here we provide arguments that alternative explanations for our data are less likely.

High SFR/CO, {\cii}/CO and {\hi}/CO ratios are usually found at low metallicities \citep[$1/6$--$1/5$ solar or below;][]{poglitsch95,israel96,madden97,madden00,rubin09,cormier10,hunt14b,hunt15,amorin16}, so the {\grb} host in principle might be a normal metal-poor dwarf. 
However,  its average metallicity is $\metoh\sim8.6$ or $\sim0.8$ solar \citep{sollerman05}, so it is difficult to advocate that its properties are due to metallicity effects.  Indeed, as shown in the previous section, the {\grb} host  has a low molecular gas mass  even taking into account its metallicity.
Moreover, the {\cii}/FIR and {\oi}/FIR ratios of the {\grb} host are higher than those of  other local dwarf galaxies with similar FIR luminosity (Fig.~\ref{fig:lum}).

Another possibility is that the {\grb} host is at the end of a star-formation episode, which would imply molecular gas dissipation (hence weak CO emission) by massive stars \citep{hatsukade14,stanway15}. However, in that case the WR region, which is forming stars most intensely, would be the most metal-rich and dust-free, because dust would be destroyed together with molecular gas. On the contrary, the WR region is metal-poor \citep{christensen08} and dusty \citep{michalowski14}. Moreover, if star formation activity was going down, then the {\cii}, {\oi} and H$\alpha$ emission would decrease almost instantaneously, whereas infrared emission would need a longer time to react. This would result in a lower (or at most equal) SFR$_{\rm line}$ than SFR$_{\rm IR}$, contrary to observations (Fig.~\ref{fig:sfrvssfr}). The dust heating from older low-mass stars would make this effect even stronger.

In order to explain the {\hi} concentration close to the WR region and the disturbed {\hi} velocity field in the {\grb} host \citet{arabsalmani15b} invoked a minor merger scenario.
This is actually not much different than our atomic gas inflow
scenario, because an {\hi}-dominated dwarf galaxy merging with the host is conceptually close to what we call an infalling {\hi} cloud. However, we do not see the enhancement of {\cii} close to the position of the WR region (similar to the {\hi} concentration), so in the merger scenario the smaller galaxy needs to be relatively un-evolved, so it does not bring significant amount of carbon.

\begin{figure}
\begin{center}
\includegraphics[width=0.5\textwidth]{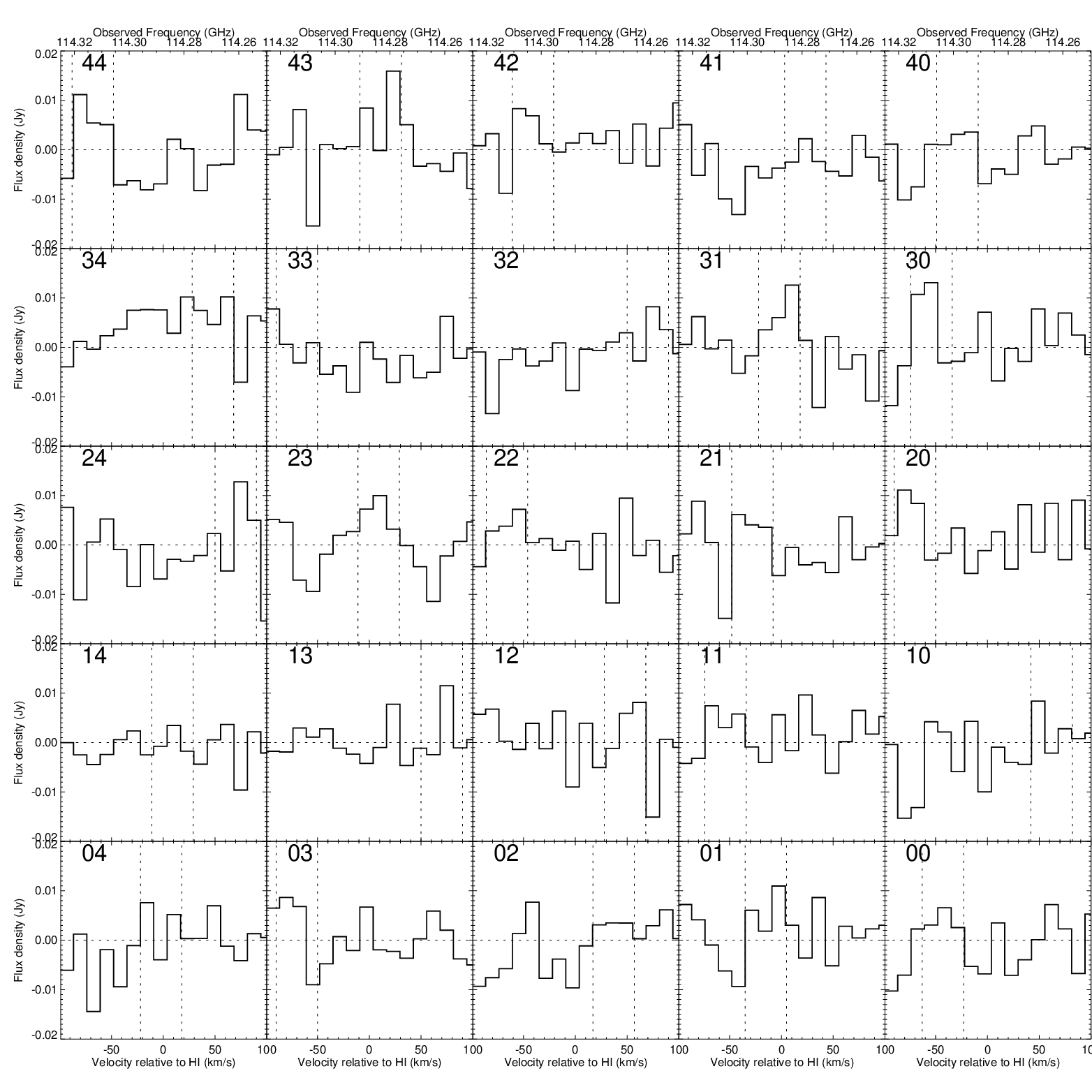} 
\end{center}
\caption{ALMA CO(1-0) spectra of all spaxels. The integrated fluxes were measured within inner dotted lines. They are different for different spaxels, because for each one we conservatively selected a $40\,\kms$ region which gives the highest flux upper limit.
}
\label{fig:cospax}
\end{figure}

\begin{figure}
\begin{center}
\includegraphics[width=0.45\textwidth]{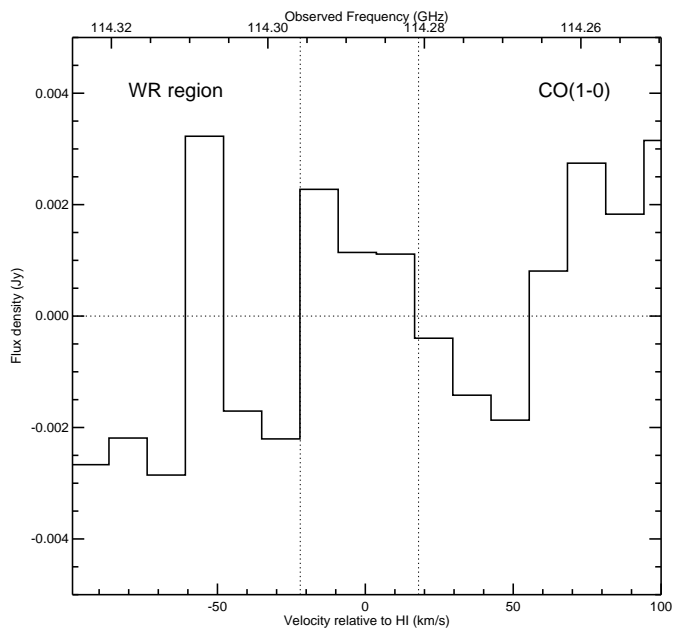} 
\end{center}
\caption{ALMA CO(1-0) spectrum of the WR region within an 1.7\arcsec\ aperture. The flux was measured within inner dotted lines.
}
\label{fig:coWR}
\end{figure}

\section{Conclusions}
\label{sec:conclusion}

Using {\cii}, {\oi} and CO spectroscopy we found that the GRB\,980425 host  has elevated {\cii}/FIR and {\oi}/FIR ratios and higher values of SFR derived from line ({\cii}, {\oi}, H$\alpha$) than from continuum (UV, IR, radio) indicators. {\cii} emission exhibits a normal morphology, peaking at the galaxy center, whereas {\oi} is concentrated close to the GRB position and the nearby Wolf-Rayet region. The high {\oi} flux indicates high radiation field and gas density at these positions, as derived from Photo Dissociation Region modelling. The {\cii}/CO luminosity ratio of the GRB\,980425 host is close to the highest values found for local star-forming galaxies. Indeed, its CO-derived molecular gas mass is low given its SFR and metallicity, but the {\cii}-derived molecular gas mass is close to the expected value.

The {\oi} and {\hi} concentrations as well as the high radiation field and density close to the GRB position are consistent with the hypothesis of a  very recent (at most a few tens of Myr ago) inflow of atomic gas triggering star formation.
In this scenario
dust has not had time to build up (explaining high line-to-continuum ratios). Such a recent enhancement of star-formation activity would indeed manifest itself in high $\mbox{SFR}_{\rm line}/\mbox{SFR}_{\rm continuum}$ ratios, because the line indicators are sensitive only to recent ($<10\,$Myr) activity, whereas the continuum indicators measure the  SFR averaged over much longer periods ($\sim100\,$Myr). We found similarly high SFR ratios for other GRB hosts. 
This is consistent with a very recent enhancement of star formation being common
among GRB hosts, so galaxies which have recently experienced  inflow of gas may preferentially host stars exploding as GRBs.
Therefore GRBs may be used to select unique samples of galaxies suitable for the investigation of recent gas accretion.

\begin{acknowledgements}

We thank Joanna Baradziej for help with improving this paper; Per Bergman, Carlos De Breuck and Katharina Immer for help with the APEX observations; Tom Muxlow, and Eelco van Kampen with the ALMA observations; and our referee for useful comments.

M.J.M.~acknowledges the support of the UK Science and Technology Facilities Council. 
The Dark Cosmology Centre is funded by the Danish National Research Foundation.
J.L.W.~is supported by a European Union COFUND/Durham Junior Research Fellowship under EU grant agreement number 267209, and acknowledges additional support from STFC (ST/L00075X/1).
A.K.~acknowledges support from the Foundation for Polish Science (FNP) and the Polish National Science Center grant 2013/11/N/ST9/00400.
A.J.C.T.~acknowledges support from the Spanish Ministry Project AYA2015-71718-R.
D.X.~acknowledges the support by the One-Hundred-Talent Program of the Chinese Academy of Sciences, and by the Strategic Priority Research Program ``Multi-wavelength Gravitational Wave Universe'' of the Chinese Academy of Sciences (No. XDB23000000).

PACS has been developed by a consortium of institutes led by MPE (Germany) and including UVIE (Austria); KU Leuven, CSL, IMEC (Belgium); CEA, LAM (France); MPIA (Germany); INAF-IFSI/OAA/OAP/OAT, LENS, SISSA (Italy); IAC (Spain). This development has been supported by the funding agencies BMVIT (Austria), ESA-PRODEX (Belgium), CEA/CNES (France), DLR (Germany), ASI/INAF (Italy), and CICYT/MCYT (Spain). 
This publication is based on data acquired with the Atacama Pathfinder Experiment (APEX). APEX is a collaboration between the Max-Planck-Institut fur Radioastronomie, the European Southern Observatory, and the Onsala Space Observatory.  
Based on observations collected at the European Organisation for Astronomical Research in the Southern Hemisphere under ESO programmes 096.D-0280 and 096.F-9302 
This paper makes use of the following ALMA data: ADS/JAO.ALMA\#2011.0.00046.S. ALMA is a partnership of ESO (representing its member states), NSF (USA) and NINS (Japan), together with NRC (Canada) and NSC and ASIAA (Taiwan), in cooperation with the Republic of Chile. The Joint ALMA Observatory is operated by ESO, AUI/NRAO and NAOJ. 
This research has made use of 
the GHostS database (\urltt{http://www.grbhosts.org}), which is partly funded by Spitzer/NASA grant RSA Agreement No. 1287913; 
the NASA/IPAC Extragalactic Database (NED) which is operated by the Jet Propulsion Laboratory, California Institute of Technology, under contract with the National Aeronautics and Space Administration;
SAOImage DS9, developed by Smithsonian Astrophysical Observatory \citep{ds9};
and the NASA's Astrophysics Data System Bibliographic Services.

\end{acknowledgements}




\begin{thebibliography}{172}
\expandafter\ifx\csname natexlab\endcsname\relax\def\natexlab#1{#1}\fi
\expandafter\ifx\csname url\endcsname\relax
  \def\url#1{{\tt #1}}\fi
\expandafter\ifx\csname urlprefix\endcsname\relax\def\urlprefix{URL }\fi

\bibitem[{{Amor{\'{\i}}n} et~al.(2016){Amor{\'{\i}}n},
  {Mu{\~n}oz-Tu{\~n}{\'o}n}, {Aguerri}, \& {Planesas}}]{amorin16}
{Amor{\'{\i}}n} R., {Mu{\~n}oz-Tu{\~n}{\'o}n} C., {Aguerri} J.A.L., {Planesas}
  P., 2016, \aap, 588, A23

\bibitem[{{Arabsalmani} et~al.(2015){Arabsalmani}, {Roychowdhury}, {Zwaan},
  {Kanekar}, \& {Micha{\l}owski}}]{arabsalmani15b}
{Arabsalmani} M., {Roychowdhury} S., {Zwaan} M.A., {Kanekar} N.,
  {Micha{\l}owski} M.J., 2015, \mnras, 454, L51

\bibitem[{{Asplund} et~al.(2004){Asplund}, {Grevesse}, {Sauval}, {Allende
  Prieto}, \& {Kiselman}}]{asplund04}
{Asplund} M., {Grevesse} N., {Sauval} A.J., {Allende Prieto} C., {Kiselman} D.,
  2004, \aap, 417, 751

\bibitem[{{Bigiel} et~al.(2008){Bigiel}, {Leroy}, {Walter} et~al.}]{bigiel08}
{Bigiel} F., {Leroy} A., {Walter} F., et~al., 2008, \aj, 136, 2846

\bibitem[{{Bigiel} et~al.(2010){Bigiel}, {Leroy}, {Walter} et~al.}]{bigiel10}
{Bigiel} F., {Leroy} A., {Walter} F., et~al., 2010, \aj, 140, 1194

\bibitem[{{Bloom} et~al.(1998){Bloom}, {Djorgovski}, {Kulkarni}, \&
  {Frail}}]{bloom}
{Bloom} J.S., {Djorgovski} S.G., {Kulkarni} S.R., {Frail} D.A., 1998, \apjl,
  507, L25

\bibitem[{{Bloom} et~al.(2001){Bloom}, {Djorgovski}, \& {Kulkarni}}]{bloom01}
{Bloom} J.S., {Djorgovski} S.G., {Kulkarni} S.R., 2001, \apj, 554, 678

\bibitem[{{Boissier} et~al.(2013){Boissier}, {Salvaterra}, {Le Floc'h}
  et~al.}]{boissier13}
{Boissier} S., {Salvaterra} R., {Le Floc'h} E., et~al., 2013, \aap, 557, A34

\bibitem[{{Bolatto} et~al.(2013){Bolatto}, {Wolfire}, \& {Leroy}}]{bolatto13}
{Bolatto} A.D., {Wolfire} M., {Leroy} A.K., 2013, \araa, 51, 207

\bibitem[{{Boselli} et~al.(2009){Boselli}, {Boissier}, {Cortese}
  et~al.}]{boselli09}
{Boselli} A., {Boissier} S., {Cortese} L., et~al., 2009, \apj, 706, 1527

\bibitem[{{Carilli} \& {Walter}(2013)}]{carilli13}
{Carilli} C.L., {Walter} F., 2013, \araa, 51, 105

\bibitem[{{Castro Cer{\'o}n} et~al.(2006){Castro Cer{\'o}n}, {Micha{\l}owski},
  {Hjorth} et~al.}]{castroceron06}
{Castro Cer{\'o}n} J.M., {Micha{\l}owski} M.J., {Hjorth} J., et~al., 2006,
  \apjl, 653, L85

\bibitem[{{Castro Cer{\'o}n} et~al.(2010){Castro Cer{\'o}n}, {Micha{\l}owski},
  {Hjorth} et~al.}]{castroceron10}
{Castro Cer{\'o}n} J.M., {Micha{\l}owski} M.J., {Hjorth} J., et~al., 2010,
  \apj, 721, 1919

\bibitem[{{Castro-Tirado} et~al.(2007){Castro-Tirado}, {Bremer}, {McBreen}
  et~al.}]{castrotirado07}
{Castro-Tirado} A.J., {Bremer} M., {McBreen} S., et~al., 2007, \aap, 475, 101

\bibitem[{{Chabrier}(2003)}]{chabrier03}
{Chabrier} G., 2003, \apjl, 586, L133

\bibitem[{{Christensen} et~al.(2004){Christensen}, {Hjorth}, \&
  {Gorosabel}}]{christensen04}
{Christensen} L., {Hjorth} J., {Gorosabel} J., 2004, \aap, 425, 913

\bibitem[{{Christensen} et~al.(2008){Christensen}, {Vreeswijk}, {Sollerman}
  et~al.}]{christensen08}
{Christensen} L., {Vreeswijk} P.M., {Sollerman} J., et~al., 2008, \aap, 490, 45

\bibitem[{{Cormier} et~al.(2010){Cormier}, {Madden}, {Hony} et~al.}]{cormier10}
{Cormier} D., {Madden} S.C., {Hony} S., et~al., 2010, \aap, 518, L57

\bibitem[{{Cormier} et~al.(2015){Cormier}, {Madden}, {Lebouteiller}
  et~al.}]{cormier15}
{Cormier} D., {Madden} S.C., {Lebouteiller} V., et~al., 2015, \aap, 578, A53

\bibitem[{{Crawford} et~al.(1985){Crawford}, {Genzel}, {Townes}, \&
  {Watson}}]{crawford85}
{Crawford} M.K., {Genzel} R., {Townes} C.H., {Watson} D.M., 1985, \apj, 291,
  755

\bibitem[{{da Silva} et~al.(2014){da Silva}, {Fumagalli}, \&
  {Krumholz}}]{dasilva14}
{da Silva} R.L., {Fumagalli} M., {Krumholz} M.R., 2014, \mnras, 444, 3275

\bibitem[{{Daddi} et~al.(2010){Daddi}, {Bournaud}, {Walter} et~al.}]{daddi10}
{Daddi} E., {Bournaud} F., {Walter} F., et~al., 2010, \apj, 713, 686

\bibitem[{{Davies} et~al.(2016){Davies}, {Driver}, {Robotham}
  et~al.}]{davies16}
{Davies} L.J.M., {Driver} S.P., {Robotham} A.S.G., et~al., 2016, \mnras, 461,
  458

\bibitem[{{De Looze} et~al.(2014){De Looze}, {Cormier}, {Lebouteiller}
  et~al.}]{delooze14}
{De Looze} I., {Cormier} D., {Lebouteiller} V., et~al., 2014, \aap, 568, A62

\bibitem[{{D'Elia} et~al.(2010){D'Elia}, {Fynbo}, {Covino} et~al.}]{delia10}
{D'Elia} V., {Fynbo} J.P.U., {Covino} S., et~al., 2010, \aap, 523, A36

\bibitem[{{D'Elia} et~al.(2014){D'Elia}, {Fynbo}, {Goldoni} et~al.}]{delia14}
{D'Elia} V., {Fynbo} J.P.U., {Goldoni} P., et~al., 2014, \aap, 564, A38

\bibitem[{{Della Valle} et~al.(2006){Della Valle}, {Chincarini}, {Panagia}
  et~al.}]{dellavalle06}
{Della Valle} M., {Chincarini} G., {Panagia} N., et~al., 2006, \nat, 444, 1050

\bibitem[{{Detmers} et~al.(2008){Detmers}, {Langer}, {Podsiadlowski}, \&
  {Izzard}}]{detmers08}
{Detmers} R.G., {Langer} N., {Podsiadlowski} P., {Izzard} R.G., 2008, \aap,
  484, 831

\bibitem[{{Djorgovski} et~al.(2001){Djorgovski}, {Frail}, {Kulkarni}
  et~al.}]{djorgovski01}
{Djorgovski} S.G., {Frail} D.A., {Kulkarni} S.R., et~al., 2001, \apj, 562, 654

\bibitem[{{Djorgovski} et~al.(2003){Djorgovski}, {Bloom}, \&
  {Kulkarni}}]{djorgovski03}
{Djorgovski} S.G., {Bloom} J.S., {Kulkarni} S.R., 2003, \apjl, 591, L13

\bibitem[{{Draine}(2009)}]{draine09}
{Draine} B.T., 2009, ASPC, 414, 453

\bibitem[{{Elliott} et~al.(2013){Elliott}, {Kr{\"u}hler}, {Greiner}
  et~al.}]{elliott13}
{Elliott} J., {Kr{\"u}hler} T., {Greiner} J., et~al., 2013, \aap, 556, A23

\bibitem[{{Elmegreen} et~al.(2016){Elmegreen}, {Kaufman}, {Bournaud}
  et~al.}]{elmegreen16}
{Elmegreen} B.G., {Kaufman} M., {Bournaud} F., et~al., 2016, \apj, 823, 26

\bibitem[{{Filho} et~al.(2016){Filho}, {S{\'a}nchez Almeida}, {Amor{\'{\i}}n}
  et~al.}]{filho16}
{Filho} M.E., {S{\'a}nchez Almeida} J., {Amor{\'{\i}}n} R., et~al., 2016, \apj,
  820, 109

\bibitem[{{Friis} et~al.(2015){Friis}, {De Cia}, {Kr{\"u}hler}
  et~al.}]{friis15}
{Friis} M., {De Cia} A., {Kr{\"u}hler} T., et~al., 2015, \mnras, 451, 4686

\bibitem[{{Fruchter} et~al.(2006){Fruchter}, {Levan}, {Strolger}
  et~al.}]{fruchter06}
{Fruchter} A.S., {Levan} A.J., {Strolger} L., et~al., 2006, \nat, 441, 463

\bibitem[{{Fumagalli} \& {Gavazzi}(2008)}]{fumagalli08}
{Fumagalli} M., {Gavazzi} G., 2008, \aap, 490, 571

\bibitem[{{Fumagalli} et~al.(2011){Fumagalli}, {da Silva}, \&
  {Krumholz}}]{fumagalli11}
{Fumagalli} M., {da Silva} R.L., {Krumholz} M.R., 2011, \apjl, 741, L26

\bibitem[{{Fynbo} et~al.(2006){Fynbo}, {Starling}, {Ledoux} et~al.}]{fynbo06b}
{Fynbo} J.P.U., {Starling} R.L.C., {Ledoux} C., et~al., 2006, \aap, 451, L47

\bibitem[{{Galama} et~al.(1998){Galama}, {Vreeswijk}, {van Paradijs}
  et~al.}]{galamanature}
{Galama} T.J., {Vreeswijk} P.M., {van Paradijs} J., et~al., 1998, \nat, 395,
  670

\bibitem[{{Garnavich} et~al.(2003){Garnavich}, {Stanek}, {Wyrzykowski}
  et~al.}]{garnavich03}
{Garnavich} P.M., {Stanek} K.Z., {Wyrzykowski} L., et~al., 2003, \apj, 582, 924

\bibitem[{{Georgy} et~al.(2012){Georgy}, {Ekstr{\"o}m}, {Meynet}
  et~al.}]{georgy12}
{Georgy} C., {Ekstr{\"o}m} S., {Meynet} G., et~al., 2012, \aap, 542, A29

\bibitem[{{Glover} \& {Clark}(2012)}]{glover12}
{Glover} S.C.O., {Clark} P.C., 2012, \mnras, 421, 9

\bibitem[{{Gorosabel} et~al.(2005){Gorosabel}, {P{\'e}rez-Ram{\'{\i}}rez},
  {Sollerman} et~al.}]{gorosabel05}
{Gorosabel} J., {P{\'e}rez-Ram{\'{\i}}rez} D., {Sollerman} J., et~al., 2005,
  \aap, 444, 711

\bibitem[{{Gullberg} et~al.(2015){Gullberg}, {De Breuck}, {Vieira}
  et~al.}]{gullberg15}
{Gullberg} B., {De Breuck} C., {Vieira} J.D., et~al., 2015, \mnras, 449, 2883

\bibitem[{{Guo} et~al.(2016){Guo}, {Rafelski}, {Faber} et~al.}]{guo16}
{Guo} Y., {Rafelski} M., {Faber} S.M., et~al., 2016, \apj, submitted ({\tt
  arXiv:1604.05314})

\bibitem[{{G{\"u}sten} et~al.(2006){G{\"u}sten}, {Nyman}, {Schilke}
  et~al.}]{apex}
{G{\"u}sten} R., {Nyman} L.{\AA}., {Schilke} P., et~al., 2006, \aap, 454, L13

\bibitem[{{Habing}(1968)}]{habing}
{Habing} H.J., 1968, \bain, 19, 421

\bibitem[{{Hammer} et~al.(2006){Hammer}, {Flores}, {Schaerer}
  et~al.}]{hammer06}
{Hammer} F., {Flores} H., {Schaerer} D., et~al., 2006, \aap, 454, 103

\bibitem[{{Han} et~al.(2010){Han}, {Hammer}, {Liang} et~al.}]{han10}
{Han} X.H., {Hammer} F., {Liang} Y.C., et~al., 2010, \aap, 514, A24

\bibitem[{{Hashimoto} et~al.(2015){Hashimoto}, {Perley}, {Ohta}
  et~al.}]{hashimoto15}
{Hashimoto} T., {Perley} D.A., {Ohta} K., et~al., 2015, \apj, 806, 250

\bibitem[{{Hatsukade} et~al.(2007){Hatsukade}, {Kohno}, {Endo}
  et~al.}]{hatsukade07}
{Hatsukade} B., {Kohno} K., {Endo} A., et~al., 2007, \pasj, 59, 67

\bibitem[{{Hatsukade} et~al.(2014){Hatsukade}, {Ohta}, {Endo}
  et~al.}]{hatsukade14}
{Hatsukade} B., {Ohta} K., {Endo} A., et~al., 2014, \nat, 510, 247

\bibitem[{{Herrera-Camus} et~al.(2015){Herrera-Camus}, {Bolatto}, {Wolfire}
  et~al.}]{herreracamus15}
{Herrera-Camus} R., {Bolatto} A.D., {Wolfire} M.G., et~al., 2015, \apj, 800, 1

\bibitem[{{Hjorth} \& {Bloom}(2012)}]{hjorthsn}
{Hjorth} J., {Bloom} J.S., 2012, Cambridge University Press, 169

\bibitem[{{Hjorth} et~al.(2003){Hjorth}, {Sollerman}, {M{\o}ller}
  et~al.}]{hjorthnature}
{Hjorth} J., {Sollerman} J., {M{\o}ller} P., et~al., 2003, \nat, 423, 847

\bibitem[{{Hjorth} et~al.(2012){Hjorth}, {Malesani}, {Jakobsson}
  et~al.}]{hjorth12}
{Hjorth} J., {Malesani} D., {Jakobsson} P., et~al., 2012, \apj, 756, 187

\bibitem[{{Hu} et~al.(2016){Hu}, {Naab}, {Walch}, {Glover}, \& {Clark}}]{hu16}
{Hu} C.Y., {Naab} T., {Walch} S., {Glover} S.C.O., {Clark} P.C., 2016, \mnras,
  458, 3528

\bibitem[{{Hughes} et~al.(2016){Hughes}, {Ibar}, {Villanueva}
  et~al.}]{hughes17}
{Hughes} T.M., {Ibar} E., {Villanueva} V., et~al., 2016, \aap, in prep.

\bibitem[{{Hunt} et~al.(2014{\natexlab{a}}){Hunt}, {Palazzi}, {Micha{\l}owski}
  et~al.}]{hunt14}
{Hunt} L.K., {Palazzi} E., {Micha{\l}owski} M.J., et~al., 2014{\natexlab{a}},
  \aap, 565, A112

\bibitem[{{Hunt} et~al.(2014{\natexlab{b}}){Hunt}, {Testi}, {Casasola}
  et~al.}]{hunt14b}
{Hunt} L.K., {Testi} L., {Casasola} V., et~al., 2014{\natexlab{b}}, \aap, 561,
  A49

\bibitem[{{Hunt} et~al.(2015){Hunt}, {Garc{\'{\i}}a-Burillo}, {Casasola}
  et~al.}]{hunt15}
{Hunt} L.K., {Garc{\'{\i}}a-Burillo} S., {Casasola} V., et~al., 2015, \aap,
  583, A114

\bibitem[{{Iglesias-P{\'a}ramo} et~al.(2004){Iglesias-P{\'a}ramo}, {Boselli},
  {Gavazzi}, \& {Zaccardo}}]{iglesias04}
{Iglesias-P{\'a}ramo} J., {Boselli} A., {Gavazzi} G., {Zaccardo} A., 2004,
  \aap, 421, 887

\bibitem[{{Iglesias-P{\'a}ramo} et~al.(2007){Iglesias-P{\'a}ramo}, {Buat},
  {Hern{\'a}ndez-Fern{\'a}ndez} et~al.}]{iglesias07}
{Iglesias-P{\'a}ramo} J., {Buat} V., {Hern{\'a}ndez-Fern{\'a}ndez} J., et~al.,
  2007, \apj, 670, 279

\bibitem[{{Israel} et~al.(1996){Israel}, {Maloney}, {Geis} et~al.}]{israel96}
{Israel} F.P., {Maloney} P.R., {Geis} N., et~al., 1996, \apj, 465, 738

\bibitem[{{Izzard} et~al.(2004){Izzard}, {Ramirez-Ruiz}, \& {Tout}}]{izzard04}
{Izzard} R.G., {Ramirez-Ruiz} E., {Tout} C.A., 2004, \mnras, 348, 1215

\bibitem[{{Jakobsson} et~al.(2012){Jakobsson}, {Hjorth}, {Malesani}
  et~al.}]{jakobsson12}
{Jakobsson} P., {Hjorth} J., {Malesani} D., et~al., 2012, \apj, 752, 62

\bibitem[{{Japelj} et~al.(2016){Japelj}, {Vergani}, {Salvaterra}
  et~al.}]{japelj16}
{Japelj} J., {Vergani} S.D., {Salvaterra} R., et~al., 2016, \aap, 590, A129

\bibitem[{{Joye} \& {Mandel}(2003)}]{ds9}
{Joye} W.A., {Mandel} E., 2003, In: {H.~E.~Payne, R.~I.~Jedrzejewski, \&
  R.~N.~Hook} (ed.) Astronomical Data Analysis Software and Systems XII, vol.
  295 of Astronomical Society of the Pacific Conference Series, 489

\bibitem[{{Kaufman} et~al.(2006){Kaufman}, {Wolfire}, \&
  {Hollenbach}}]{kaufman06}
{Kaufman} M.J., {Wolfire} M.G., {Hollenbach} D.J., 2006, \apj, 644, 283

\bibitem[{{Kennicutt}(1998)}]{kennicutt}
{Kennicutt} R.C., 1998, \araa, 36, 189

\bibitem[{{Kewley} \& {Dopita}(2002)}]{kewley02}
{Kewley} L.J., {Dopita} M.A., 2002, \apjs, 142, 35

\bibitem[{{Kewley} et~al.(2004){Kewley}, {Geller}, \& {Jansen}}]{kewley04}
{Kewley} L.J., {Geller} M.J., {Jansen} R.A., 2004, \aj, 127, 2002

\bibitem[{{Kr{\"u}hler} et~al.(2012){Kr{\"u}hler}, {Fynbo}, {Geier}
  et~al.}]{kruhler12}
{Kr{\"u}hler} T., {Fynbo} J.P.U., {Geier} S., et~al., 2012, \aap, 546, A8

\bibitem[{{Kr{\"u}hler} et~al.(2013){Kr{\"u}hler}, {Ledoux}, {Fynbo}
  et~al.}]{kruhler13}
{Kr{\"u}hler} T., {Ledoux} C., {Fynbo} J.P.U., et~al., 2013, \aap, 557, A18

\bibitem[{{Krumholz}(2012)}]{krumholz12}
{Krumholz} M.R., 2012, \apj, 759, 9

\bibitem[{{Le Floc'h} et~al.(2006){Le Floc'h}, {Charmandaris}, {Forrest}
  et~al.}]{lefloch}
{Le Floc'h} E., {Charmandaris} V., {Forrest} W.J., et~al., 2006, \apj, 642, 636

\bibitem[{{Lee} et~al.(2009){Lee}, {Gil de Paz}, {Tremonti} et~al.}]{lee09b}
{Lee} J.C., {Gil de Paz} A., {Tremonti} C., et~al., 2009, \apj, 706, 599

\bibitem[{{Levesque} et~al.(2010{\natexlab{a}}){Levesque}, {Berger}, {Kewley},
  \& {Bagley}}]{levesque10c}
{Levesque} E.M., {Berger} E., {Kewley} L.J., {Bagley} M.M., 2010{\natexlab{a}},
  \aj, 139, 694

\bibitem[{{Levesque} et~al.(2010{\natexlab{b}}){Levesque}, {Kewley}, {Graham},
  \& {Fruchter}}]{levesque10b}
{Levesque} E.M., {Kewley} L.J., {Graham} J.F., {Fruchter} A.S.,
  2010{\natexlab{b}}, \apjl, 712, L26

\bibitem[{{Madden}(2000)}]{madden00}
{Madden} S.C., 2000, \nar, 44, 249

\bibitem[{{Madden} et~al.(1997){Madden}, {Poglitsch}, {Geis}, {Stacey}, \&
  {Townes}}]{madden97}
{Madden} S.C., {Poglitsch} A., {Geis} N., {Stacey} G.J., {Townes} C.H., 1997,
  \apj, 483, 200

\bibitem[{{Madden} et~al.(2013){Madden}, {R{\'e}my-Ruyer}, {Galametz}
  et~al.}]{madden13}
{Madden} S.C., {R{\'e}my-Ruyer} A., {Galametz} M., et~al., 2013, \pasp, 125,
  600

\bibitem[{{Maiolino} et~al.(2008){Maiolino}, {Nagao}, {Grazian}
  et~al.}]{maiolino08}
{Maiolino} R., {Nagao} T., {Grazian} A., et~al., 2008, \aap, 488, 463

\bibitem[{{Malhotra} et~al.(2001){Malhotra}, {Kaufman}, {Hollenbach}
  et~al.}]{malhotra01}
{Malhotra} S., {Kaufman} M.J., {Hollenbach} D., et~al., 2001, \apj, 561, 766

\bibitem[{{Martin} et~al.(2014){Martin}, {Chang}, {Matuszewski}
  et~al.}]{martin14}
{Martin} D.C., {Chang} D., {Matuszewski} M., et~al., 2014, \apj, 786, 106

\bibitem[{{McMullin} et~al.(2007){McMullin}, {Waters}, {Schiebel}, {Young}, \&
  {Golap}}]{casa}
{McMullin} J.P., {Waters} B., {Schiebel} D., {Young} W., {Golap} K., 2007, In:
  {Shaw} R.A., {Hill} F., {Bell} D.J. (eds.) Astronomical Data Analysis
  Software and Systems XVI, vol. 376 of Astronomical Society of the Pacific
  Conference Series, 127

\bibitem[{{Meurer} et~al.(2009){Meurer}, {Wong}, {Kim} et~al.}]{meurer09}
{Meurer} G.R., {Wong} O.I., {Kim} J.H., et~al., 2009, \apj, 695, 765

\bibitem[{{Micha{\l}owski} et~al.(2008){Micha{\l}owski}, {Hjorth}, {Castro
  Cer{\'o}n}, \& {Watson}}]{michalowski08}
{Micha{\l}owski} M.J., {Hjorth} J., {Castro Cer{\'o}n} J.M., {Watson} D., 2008,
  \apj, 672, 817

\bibitem[{{Micha{\l}owski} et~al.(2009){Micha{\l}owski}, {Hjorth}, {Malesani}
  et~al.}]{michalowski09}
{Micha{\l}owski} M.J., {Hjorth} J., {Malesani} D., et~al., 2009, \apj, 693, 347

\bibitem[{{Micha{\l}owski} et~al.(2010{\natexlab{a}}){Micha{\l}owski},
  {Hjorth}, \& {Watson}}]{michalowski10smg}
{Micha{\l}owski} M.J., {Hjorth} J., {Watson} D., 2010{\natexlab{a}}, \aap, 514,
  A67

\bibitem[{{Micha{\l}owski} et~al.(2010{\natexlab{b}}){Micha{\l}owski},
  {Watson}, \& {Hjorth}}]{michalowski10smg4}
{Micha{\l}owski} M.J., {Watson} D., {Hjorth} J., 2010{\natexlab{b}}, \apj, 712,
  942

\bibitem[{{Micha{\l}owski} et~al.(2012{\natexlab{a}}){Micha{\l}owski},
  {Dunlop}, {Cirasuolo} et~al.}]{michalowski12mass}
{Micha{\l}owski} M.J., {Dunlop} J.S., {Cirasuolo} M., et~al.,
  2012{\natexlab{a}}, \aap, 541, A85

\bibitem[{{Micha{\l}owski} et~al.(2012{\natexlab{b}}){Micha{\l}owski},
  {Kamble}, {Hjorth} et~al.}]{michalowski12grb}
{Micha{\l}owski} M.J., {Kamble} A., {Hjorth} J., et~al., 2012{\natexlab{b}},
  \apj, 755, 85

\bibitem[{{Micha{\l}owski} et~al.(2014{\natexlab{a}}){Micha{\l}owski},
  {Hayward}, {Dunlop} et~al.}]{michalowski14mass}
{Micha{\l}owski} M.J., {Hayward} C.C., {Dunlop} J.S., et~al.,
  2014{\natexlab{a}}, \aap, 571, A75

\bibitem[{{Micha{\l}owski} et~al.(2014{\natexlab{b}}){Micha{\l}owski}, {Hunt},
  {Palazzi} et~al.}]{michalowski14}
{Micha{\l}owski} M.J., {Hunt} L.K., {Palazzi} E., et~al., 2014{\natexlab{b}},
  \aap, 562, A70

\bibitem[{{Micha{\l}owski} et~al.(2015){Micha{\l}owski}, {Gentile}, {Hjorth}
  et~al.}]{michalowski15hi}
{Micha{\l}owski} M.J., {Gentile} G., {Hjorth} J., et~al., 2015, \aap, 582, A78

\bibitem[{{Mitchell} et~al.(2013){Mitchell}, {Lacey}, {Baugh}, \&
  {Cole}}]{mitchell13}
{Mitchell} P.D., {Lacey} C.G., {Baugh} C.M., {Cole} S., 2013, \mnras, 435, 87

\bibitem[{{Modjaz} et~al.(2008){Modjaz}, {Kewley}, {Kirshner}
  et~al.}]{modjaz08}
{Modjaz} M., {Kewley} L., {Kirshner} R.P., et~al., 2008, \aj, 135, 1136

\bibitem[{{Murphy} et~al.(2011){Murphy}, {Condon}, {Schinnerer}
  et~al.}]{murphy11}
{Murphy} E.J., {Condon} J.J., {Schinnerer} E., et~al., 2011, \apj, 737, 67

\bibitem[{{Neri} et~al.(2014){Neri}, {Downes}, {Cox}, \& {Walter}}]{neri14}
{Neri} R., {Downes} D., {Cox} P., {Walter} F., 2014, \aap, 562, A35

\bibitem[{{{\"O}stlin} et~al.(2008){{\"O}stlin}, {Zackrisson}, {Sollerman},
  {Mattila}, \& {Hayes}}]{ostlin08}
{{\"O}stlin} G., {Zackrisson} E., {Sollerman} J., {Mattila} S., {Hayes} M.,
  2008, \mnras, 387, 1227

\bibitem[{{Ott}(2010)}]{hipe}
{Ott} S., 2010, In: {Mizumoto} Y., {Morita} K.I., {Ohishi} M. (eds.)
  Astronomical Data Analysis Software and Systems XIX, vol. 434 of Astronomical
  Society of the Pacific Conference Series, 139

\bibitem[{{Pacifici} et~al.(2012){Pacifici}, {Charlot}, {Blaizot}, \&
  {Brinchmann}}]{pacifici12}
{Pacifici} C., {Charlot} S., {Blaizot} J., {Brinchmann} J., 2012, \mnras, 421,
  2002

\bibitem[{{Pagel} et~al.(1979){Pagel}, {Edmunds}, {Blackwell}, {Chun}, \&
  {Smith}}]{pagel79}
{Pagel} B.E.J., {Edmunds} M.G., {Blackwell} D.E., {Chun} M.S., {Smith} G.,
  1979, \mnras, 189, 95

\bibitem[{{Perley} \& {Perley}(2013)}]{perley13b}
{Perley} D.A., {Perley} R.A., 2013, \apj, 778, 172

\bibitem[{{Perley} et~al.(2013){Perley}, {Levan}, {Tanvir} et~al.}]{perley13}
{Perley} D.A., {Levan} A.J., {Tanvir} N.R., et~al., 2013, \apj, 778, 128

\bibitem[{{Perley} et~al.(2015){Perley}, {Perley}, {Hjorth} et~al.}]{perley15}
{Perley} D.A., {Perley} R.A., {Hjorth} J., et~al., 2015, \apj, 801, 102

\bibitem[{{Perley} et~al.(2016){Perley}, {Tanvir}, {Hjorth} et~al.}]{perley16b}
{Perley} D.A., {Tanvir} N.R., {Hjorth} J., et~al., 2016, \apj, 817, 8

\bibitem[{{Petrovic} et~al.(2005){Petrovic}, {Langer}, {Yoon}, \&
  {Heger}}]{petrovic05}
{Petrovic} J., {Langer} N., {Yoon} S.C., {Heger} A., 2005, \aap, 435, 247

\bibitem[{{Pettini} \& {Pagel}(2004)}]{pettini04}
{Pettini} M., {Pagel} B.E.J., 2004, \mnras, 348, L59

\bibitem[{{Pety}(2005)}]{gildas}
{Pety} J., 2005, In: {Casoli} F., {Contini} T., {Hameury} J.M., {Pagani} L.
  (eds.) SF2A-2005: Semaine de l'Astrophysique Francaise, 721

\bibitem[{{Pilbratt} et~al.(2010){Pilbratt}, {Riedinger}, {Passvogel}
  et~al.}]{herschel}
{Pilbratt} G.L., {Riedinger} J.R., {Passvogel} T., et~al., 2010, \aap, 518, L1

\bibitem[{{Piranomonte} et~al.(2015){Piranomonte}, {Japelj}, {Vergani}
  et~al.}]{piranomonte15}
{Piranomonte} S., {Japelj} J., {Vergani} S.D., et~al., 2015, \mnras, 452, 3293

\bibitem[{{Podsiadlowski} et~al.(2004){Podsiadlowski}, {Mazzali}, {Nomoto},
  {Lazzati}, \& {Cappellaro}}]{podsiadlowski04}
{Podsiadlowski} P., {Mazzali} P.A., {Nomoto} K., {Lazzati} D., {Cappellaro} E.,
  2004, \apjl, 607, L17

\bibitem[{{Podsiadlowski} et~al.(2010){Podsiadlowski}, {Ivanova}, {Justham}, \&
  {Rappaport}}]{podsiadlowski10}
{Podsiadlowski} P., {Ivanova} N., {Justham} S., {Rappaport} S., 2010, \mnras,
  406, 840

\bibitem[{{Poglitsch} et~al.(1995){Poglitsch}, {Krabbe}, {Madden}
  et~al.}]{poglitsch95}
{Poglitsch} A., {Krabbe} A., {Madden} S.C., et~al., 1995, \apj, 454, 293

\bibitem[{{Poglitsch} et~al.(2010){Poglitsch}, {Waelkens}, {Geis}
  et~al.}]{pacs}
{Poglitsch} A., {Waelkens} C., {Geis} N., et~al., 2010, \aap, 518, L2

\bibitem[{{Pound} \& {Wolfire}(2008)}]{pound08}
{Pound} M.W., {Wolfire} M.G., 2008, In: {Argyle} R.W., {Bunclark} P.S., {Lewis}
  J.R. (eds.) Astronomical Data Analysis Software and Systems XVII, vol. 394 of
  Astronomical Society of the Pacific Conference Series, 654

\bibitem[{{Price} et~al.(2002){Price}, {Kulkarni}, {Berger} et~al.}]{price02}
{Price} P.A., {Kulkarni} S.R., {Berger} E., et~al., 2002, \apjl, 571, L121

\bibitem[{{Prochaska} et~al.(2004){Prochaska}, {Bloom}, {Chen}
  et~al.}]{prochaska04}
{Prochaska} J.X., {Bloom} J.S., {Chen} H.W., et~al., 2004, \apj, 611, 200

\bibitem[{{Prochaska} et~al.(2009){Prochaska}, {Sheffer}, {Perley}
  et~al.}]{prochaska09}
{Prochaska} J.X., {Sheffer} Y., {Perley} D.A., et~al., 2009, \apjl, 691, L27

\bibitem[{{Rafelski} et~al.(2016){Rafelski}, {Gardner}, {Fumagalli}
  et~al.}]{rafelski16}
{Rafelski} M., {Gardner} J.P., {Fumagalli} M., et~al., 2016, \apj, 825, 87

\bibitem[{{Rauch} et~al.(2016){Rauch}, {Becker}, \& {Haehnelt}}]{rauch16}
{Rauch} M., {Becker} G.D., {Haehnelt} M.G., 2016, \mnras, 455, 3991

\bibitem[{{Ribaudo} et~al.(2011){Ribaudo}, {Lehner}, {Howk} et~al.}]{ribaudo11}
{Ribaudo} J., {Lehner} N., {Howk} J.C., et~al., 2011, \apj, 743, 207

\bibitem[{{Rigopoulou} et~al.(2014){Rigopoulou}, {Hopwood}, {Magdis}
  et~al.}]{rigopoulou14}
{Rigopoulou} D., {Hopwood} R., {Magdis} G.E., et~al., 2014, \apjl, 781, L15

\bibitem[{{Rubin} et~al.(2009){Rubin}, {Hony}, {Madden} et~al.}]{rubin09}
{Rubin} D., {Hony} S., {Madden} S.C., et~al., 2009, \aap, 494, 647

\bibitem[{{Salpeter}(1955)}]{salpeter}
{Salpeter} E.E., 1955, \apj, 121, 161

\bibitem[{{S{\'a}nchez Almeida} et~al.(2013){S{\'a}nchez Almeida},
  {Mu{\~n}oz-Tu{\~n}{\'o}n}, {Elmegreen}, {Elmegreen}, \&
  {M{\'e}ndez-Abreu}}]{sanchezalmeida13}
{S{\'a}nchez Almeida} J., {Mu{\~n}oz-Tu{\~n}{\'o}n} C., {Elmegreen} D.M.,
  {Elmegreen} B.G., {M{\'e}ndez-Abreu} J., 2013, \apj, 767, 74

\bibitem[{{S{\'a}nchez Almeida} et~al.(2014{\natexlab{a}}){S{\'a}nchez
  Almeida}, {Elmegreen}, {Mu{\~n}oz-Tu{\~n}{\'o}n}, \&
  {Elmegreen}}]{sanchezalmeida14b}
{S{\'a}nchez Almeida} J., {Elmegreen} B.G., {Mu{\~n}oz-Tu{\~n}{\'o}n} C.,
  {Elmegreen} D.M., 2014{\natexlab{a}}, \aapr, 22, 71

\bibitem[{{S{\'a}nchez Almeida} et~al.(2014{\natexlab{b}}){S{\'a}nchez
  Almeida}, {Morales-Luis}, {Mu{\~n}oz-Tu{\~n}{\'o}n}
  et~al.}]{sanchezalmeida14}
{S{\'a}nchez Almeida} J., {Morales-Luis} A.B., {Mu{\~n}oz-Tu{\~n}{\'o}n} C.,
  et~al., 2014{\natexlab{b}}, \apj, 783, 45

\bibitem[{{Sancisi} et~al.(2008){Sancisi}, {Fraternali}, {Oosterloo}, \& {van
  der Hulst}}]{sancisi08}
{Sancisi} R., {Fraternali} F., {Oosterloo} T., {van der Hulst} T., 2008, \aapr,
  15, 189

\bibitem[{{Sargsyan} et~al.(2012){Sargsyan}, {Lebouteiller}, {Weedman}
  et~al.}]{sargsyan12}
{Sargsyan} L., {Lebouteiller} V., {Weedman} D., et~al., 2012, \apj, 755, 171

\bibitem[{{Savaglio} et~al.(2009){Savaglio}, {Glazebrook}, \&
  {LeBorgne}}]{savaglio09}
{Savaglio} S., {Glazebrook} K., {LeBorgne} D., 2009, \apj, 691, 182

\bibitem[{{Savaglio} et~al.(2012){Savaglio}, {Rau}, {Greiner}
  et~al.}]{savaglio12}
{Savaglio} S., {Rau} A., {Greiner} J., et~al., 2012, \mnras, 420, 627

\bibitem[{{Schady} et~al.(2014){Schady}, {Savaglio}, {M{\"u}ller}
  et~al.}]{schady14}
{Schady} P., {Savaglio} S., {M{\"u}ller} T., et~al., 2014, \aap, 570, A52

\bibitem[{{Schady} et~al.(2015){Schady}, {Kr{\"u}hler}, {Greiner}
  et~al.}]{schady15}
{Schady} P., {Kr{\"u}hler} T., {Greiner} J., et~al., 2015, \aap, 579, A126

\bibitem[{{Schaye} et~al.(2010){Schaye}, {Dalla Vecchia}, {Booth}
  et~al.}]{schaye10}
{Schaye} J., {Dalla Vecchia} C., {Booth} C.M., et~al., 2010, \mnras, 402, 1536

\bibitem[{{Schulze} et~al.(2014){Schulze}, {Malesani}, {Cucchiara}
  et~al.}]{schulze14}
{Schulze} S., {Malesani} D., {Cucchiara} A., et~al., 2014, \aap, 566, A102

\bibitem[{{Schulze} et~al.(2015){Schulze}, {Chapman}, {Hjorth}
  et~al.}]{schulze15}
{Schulze} S., {Chapman} R., {Hjorth} J., et~al., 2015, \apj, 808, 73

\bibitem[{{Silva} et~al.(1998){Silva}, {Granato}, {Bressan}, \&
  {Danese}}]{silva98}
{Silva} L., {Granato} G.L., {Bressan} A., {Danese} L., 1998, \apj, 509, 103

\bibitem[{{Simha} et~al.(2014){Simha}, {Weinberg}, {Conroy} et~al.}]{simha14}
{Simha} V., {Weinberg} D.H., {Conroy} C., et~al., 2014, \mnras, submitted ({\tt
  arXiv:1404.0402})

\bibitem[{{Sollerman} et~al.(2005){Sollerman}, {{\"O}stlin}, {Fynbo}
  et~al.}]{sollerman05}
{Sollerman} J., {{\"O}stlin} G., {Fynbo} J.P.U., et~al., 2005, New Astronomy,
  11, 103

\bibitem[{{Sollerman} et~al.(2006){Sollerman}, {Jaunsen}, {Fynbo}
  et~al.}]{sollerman06}
{Sollerman} J., {Jaunsen} A.O., {Fynbo} J.P.U., et~al., 2006, \aap, 454, 503

\bibitem[{{Solomon} et~al.(1997){Solomon}, {Downes}, {Radford}, \&
  {Barrett}}]{solomon97}
{Solomon} P.M., {Downes} D., {Radford} S.J.E., {Barrett} J.W., 1997, \apj, 478,
  144

\bibitem[{{Stacey} et~al.(1991){Stacey}, {Geis}, {Genzel} et~al.}]{stacey91}
{Stacey} G.J., {Geis} N., {Genzel} R., et~al., 1991, \apj, 373, 423

\bibitem[{{Stacey} et~al.(2010){Stacey}, {Hailey-Dunsheath}, {Ferkinhoff}
  et~al.}]{stacey10}
{Stacey} G.J., {Hailey-Dunsheath} S., {Ferkinhoff} C., et~al., 2010, \apj, 724,
  957

\bibitem[{{Stanek} et~al.(2003){Stanek}, {Matheson}, {Garnavich}
  et~al.}]{stanek}
{Stanek} K.Z., {Matheson} T., {Garnavich} P.M., et~al., 2003, \apjl, 591, L17

\bibitem[{{Stanway} et~al.(2010){Stanway}, {Davies}, \& {Levan}}]{stanway10}
{Stanway} E.R., {Davies} L.J.M., {Levan} A.J., 2010, \mnras, 409, L74

\bibitem[{{Stanway} et~al.(2015{\natexlab{a}}){Stanway}, {Levan}, {Tanvir}
  et~al.}]{stanway15b}
{Stanway} E.R., {Levan} A.J., {Tanvir} N., et~al., 2015{\natexlab{a}}, \mnras,
  446, 3911

\bibitem[{{Stanway} et~al.(2015{\natexlab{b}}){Stanway}, {Levan}, {Tanvir},
  {Wiersema}, \& {van der Laan}}]{stanway15}
{Stanway} E.R., {Levan} A.J., {Tanvir} N.R., {Wiersema} K., {van der Laan}
  T.P.R., 2015{\natexlab{b}}, \apjl, 798, L7

\bibitem[{{Stott} et~al.(2013){Stott}, {Sobral}, {Bower} et~al.}]{stott13}
{Stott} J.P., {Sobral} D., {Bower} R., et~al., 2013, \mnras, 436, 1130

\bibitem[{{Sullivan} et~al.(2000){Sullivan}, {Treyer}, {Ellis}
  et~al.}]{sullivan00}
{Sullivan} M., {Treyer} M.A., {Ellis} R.S., et~al., 2000, \mnras, 312, 442

\bibitem[{{Swinbank} et~al.(2012){Swinbank}, {Karim}, {Smail}
  et~al.}]{swinbank12}
{Swinbank} A.M., {Karim} A., {Smail} I., et~al., 2012, \mnras, 427, 1066

\bibitem[{{Th{\"o}ne} et~al.(2008){Th{\"o}ne}, {Fynbo}, {{\"O}stlin}
  et~al.}]{thone08}
{Th{\"o}ne} C.C., {Fynbo} J.P.U., {{\"O}stlin} G., et~al., 2008, \apj, 676,
  1151

\bibitem[{{Tinney} et~al.(1998){Tinney}, {Stathakis}, {Cannon}
  et~al.}]{tinney98}
{Tinney} C., {Stathakis} R., {Cannon} R., et~al., 1998, \iaucirc, 6896

\bibitem[{{Trenti} et~al.(2015){Trenti}, {Perna}, \& {Jimenez}}]{trenti15}
{Trenti} M., {Perna} R., {Jimenez} R., 2015, \apj, 802, 103

\bibitem[{{Tumlinson} et~al.(2007){Tumlinson}, {Prochaska}, {Chen},
  {Dessauges-Zavadsky}, \& {Bloom}}]{tumlinson07}
{Tumlinson} J., {Prochaska} J.X., {Chen} H.W., {Dessauges-Zavadsky} M., {Bloom}
  J.S., 2007, \apj, 668, 667

\bibitem[{{Turner} et~al.(2015){Turner}, {Beck}, {Benford} et~al.}]{turner15}
{Turner} J.L., {Beck} S.C., {Benford} D.J., et~al., 2015, \nat, 519, 331

\bibitem[{{van den Heuvel} \& {Portegies Zwart}(2013)}]{vandenheuvel13}
{van den Heuvel} E.P.J., {Portegies Zwart} S.F., 2013, \apj, 779, 114

\bibitem[{{Vassilev} et~al.(2008){Vassilev}, {Meledin}, {Lapkin}
  et~al.}]{shefi}
{Vassilev} V., {Meledin} D., {Lapkin} I., et~al., 2008, \aap, 490, 1157

\bibitem[{{Vergani} et~al.(2015){Vergani}, {Salvaterra}, {Japelj}
  et~al.}]{vergani15}
{Vergani} S.D., {Salvaterra} R., {Japelj} J., et~al., 2015, \aap, 581, A102

\bibitem[{{Vreeswijk} et~al.(2004){Vreeswijk}, {Ellison}, {Ledoux}
  et~al.}]{vreeswijk04}
{Vreeswijk} P.M., {Ellison} S.L., {Ledoux} C., et~al., 2004, \aap, 419, 927

\bibitem[{{Wang} et~al.(2015){Wang}, {Serra}, {J{\'o}zsa} et~al.}]{wang15}
{Wang} J., {Serra} P., {J{\'o}zsa} G.I.G., et~al., 2015, \mnras, 453, 2399

\bibitem[{{Wang} et~al.(2016){Wang}, {Norberg}, {Gunawardhana} et~al.}]{wang16}
{Wang} L., {Norberg} P., {Gunawardhana} M.L.P., et~al., 2016, \mnras, 461, 1898

\bibitem[{{Watson} et~al.(2011){Watson}, {French}, {Christensen}
  et~al.}]{watson11}
{Watson} D., {French} J., {Christensen} L., et~al., 2011, \apj, 741, 58

\bibitem[{{Weisz} et~al.(2012){Weisz}, {Johnson}, {Johnson} et~al.}]{weisz12}
{Weisz} D.R., {Johnson} B.D., {Johnson} L.C., et~al., 2012, \apj, 744, 44

\bibitem[{{Wijesinghe} et~al.(2011){Wijesinghe}, {da Cunha}, {Hopkins}
  et~al.}]{wijesinghe11}
{Wijesinghe} D.B., {da Cunha} E., {Hopkins} A.M., et~al., 2011, \mnras, 415,
  1002

\bibitem[{{Wolfire} et~al.(1989){Wolfire}, {Hollenbach}, \&
  {Tielens}}]{wolfire89}
{Wolfire} M.G., {Hollenbach} D., {Tielens} A.G.G.M., 1989, \apj, 344, 770

\bibitem[{{Woosley} \& {Heger}(2006)}]{woosley06}
{Woosley} S.E., {Heger} A., 2006, \apj, 637, 914

\bibitem[{{Yoon} \& {Langer}(2005)}]{yoon05}
{Yoon} S.C., {Langer} N., 2005, \aap, 443, 643

\bibitem[{{Yoon} et~al.(2006){Yoon}, {Langer}, \& {Norman}}]{yoon06}
{Yoon} S.C., {Langer} N., {Norman} C., 2006, \aap, 460, 199

\end{thebibliography}
\end{document}